\documentclass[aps,prb,amsmath,amssymb]{revtex4-2}
\usepackage[latin9]{inputenc}
\usepackage{color}
\usepackage{amsmath}
\usepackage{amssymb}
\usepackage{graphicx}
\usepackage[unicode=true,
 bookmarks=true,bookmarksnumbered=false,bookmarksopen=false,
 breaklinks=false,pdfborder={0 0 1},backref=false,colorlinks=true]
 {hyperref}
\hypersetup{
 colorlinks}

\makeatletter

\providecommand{\tabularnewline}{\\}

\usepackage[varg]{txfonts}
\usepackage{tikz}
\usetikzlibrary{shapes}
\usepackage{xcolor}

\usepackage{bm}

\newcommand{\bigDiamond}{\mathop{\mathpalette\bigDi@mond\relax}}
\newcommand{\bigDi@mond}[2]{\vcenter{\hbox{\m@th \scalebox{\ifx#1\displaystyle 2\else1.2\fi}{$#1\Diamond$}}}}
\newcommand{\RNum}[1]{\uppercase\expandafter{\romannumeral #1\relax}}

\def\XXint#1#2#3{{\setbox0=\hbox{$#1{#2#3}{\int}$}
		\vcenter{\hbox{$#2#3$}}\kern-.5\wd0}}

\def\be{\begin{equation}}
	\def\ee{\end{equation}}

\def\bi{\begin{itemize}}
	\def\ei{\end{itemize}}
\def\bn{\begin{enumerate}}
	\def\en{\end{enumerate}}
\def\bea{\begin{eqnarray}}
	\def\eea{\end{eqnarray}}
\newcommand{\bpm}{\begin{pmatrix}}
	\newcommand{\epm}{\end{pmatrix}}

\def\ba{\begin{array}}
	\def\ea{\end{array}}
\def\bd{\begin{displaymath}}
	\def\ed{\end{displaymath}}

\renewcommand{\imath}{\hspace{1pt}\mathrm{i}\hspace{1pt}}

\makeatother

\begin{document}
\title{Spin liquid phase in the Hubbard model: Luttinger-Ward analysis of
the slave-rotor formalism}
\author{Xia-Ming Zheng}
\affiliation{Department of Physics, Sharif University of Technology, Tehran 14588-89694,
Iran}
\author{Mehdi Kargarian}
\email{kargarian@sharif.edu}

\affiliation{Department of Physics, Sharif University of Technology, Tehran 14588-89694,
Iran}
\begin{abstract}
We propose an approach for studying the spin liquid phase of the Hubbard
model on the triangular lattice by combining the Baym--Kadanoff formalism
with the slave rotor parton construction. This method enables the
computation of a series of two-body Feynman diagrams for the Luttinger--Ward
(LW) functional using a one-loop truncation. This approach enables
us to study the U(1) quantum spin liquid phase characterized by a
spinon Fermi surface and to derive the Green's functions for spinons,
chargons, and electrons. Our findings extend beyond the standard mean-field
approximation by accounting for the effects of gauge field fluctuations.The
spatial components of the U(1) gauge field are equivalently represented
by interactions that incorporate corrections from the spinon-chargon
two-particle random phase approximation. This framework effectively
captures the long-range correlations inherent to the U(1) quantum
spin liquid and combines non-perturbative quantum field theory with
the projective construction, providing new insights into the study
of quantum spin liquids and other strongly correlated electron systems.
We demonstrate that our approach correctly computes the low-temperature
linear temperature dependence of the specific heat in the U(1) spin
liquid, in agreement with the behavior expected for a Fermi surface.
Moreover, this approach reproduces the resonant peaks in the Mott
gap, as observed in cobalt atoms on single-layer 1$T$-TaSe$_{2}$. 
\end{abstract}
\maketitle

\section{Introduction}

In tackling the challenges posed by strong correlation problems in
quantum many-body systems, perturbation theories based on free-particle
ground states often prove inadequate. Nevertheless, in many situations,
non-perturbative effects are crucial for accurately capturing the
behavior of such systems. A commonly employed approach entails solving
Green's functions using non-perturbative quantum field theory techniques.
Within this framework, a quintessential example is the approximate
computation of the Schwinger-Dyson equations, where suitable truncations
are introduced to yield relatively precise results. Additionally,
self-consistent methods like GW theory \citep{Quinn1958,Hedin1965,Aryasetiawan1998}
and the Luttinger--Ward (LW) functional theory \citep{Luttinger1960,Baym1961,Baym1962,Dupuis2023,Stefanucci2025}
(also known as the two-particle effective action theory, $\Phi$-derivable
theory, or Baym--Kadanoff (BK) theory) have witnessed remarkable
advancements in both theoretical developments and practical applications
in recent years \citep{Li2024,Sun2021,Li2023,Sumita2023,Witt2021}.
However, due to the inclusion of infinite-order Feynman diagrams in
vertex functions, exact solutions remain unattainable. The computation
of higher-order vertex corrections presents monumental challenges
in terms of computational complexity, making it generally infeasible
to calculate high-loop Green's functions on large-scale lattices.

Beyond directly computing non-perturbative many-body Green's functions,
an alternative strategy is to manually incorporate non-perturbative
many-body correlation effects. A typical scheme in this vein is the
projective construction, encompassing formulations such as Abrikosov
fermion \citep{Abrikosov1965}, Schwinger boson \citep{Schwinger2015},
slave rotor \citep{Florens2002,Florens2004}, auxiliary spin \citep{Sachdev2023,Christos2023}
etc. The central idea is to transform the original Hamiltonian to
manifest as fractionalized quasiparticles rather than the real particles
of the original Hamiltonian. This method has successfully elucidated
phenomena like the Kondo effect \citep{Sachdev2023,Hewson1997,Coleman2015}
and heavy fermion behavior \citep{Burdin2002,Coleman1983,Hewson1997,Coleman2015}
and has qualitatively described Mott transitions and spin liquids
\citep{Florens2002,Florens2004,Law2017,Lee2006,Podolsky2009,Senthil2008}.
However, there are debates regarding quantitative calculations \citep{Lee2006,Hermele2004,Lee2008,Zhang2020,Christos2023},
especially at intermediate interaction strengths. Notably, recent
works reveal that to accurately describe the system at the intermediate
interaction strengths, one must transcend the mean-field approximation
accounting for bound states of fractionalized quasiparticles and gauge
fluctuations around saddle points \citep{Podolsky2009,Lee2005,Lee2009,He2022,Chen2022}.
Taking the potential ground state of a U(1) quantum spin liquid (QSL)
with a spinon Fermi surface (SFS) in the triangular lattice Hubbard
model as an example, increasing interaction strength drives the system
from a Fermi liquid phase to a Mott insulator phase with spin liquid
ground state. Many-body correlation effects cause electrons to fractionalize
into distinct excitations known as spinons and chargons. Spinons carry
spin-$1/2$ degrees of freedom and are electrically neutral fermions,
while chargons are complex scalar fields with charge $-e$. In the
spin liquid phase, chargons exhibit an energy gap, while spinons form
a Fermi surface. Furthermore, gauge fluctuations within the system
induce novel physical effects beyond the mean-field level, such as
spinon Kondo effects, collective spin wave excitations and so on \citep{Chen2022,He2022,Zheng2024,Balents2020}.

In this paper, we present a novel method for addressing the Hubbard
model using the slave-rotor formulation. Our method integrates techniques
from quantum field theory with parton constructions. Specifically,
we employ the Luttinger--Ward (LW) functional to compute the Green's
functions of various quantum fields, thereby enabling us to extract
both the single-particle and two-particle spectral properties as well
as the system's transport characteristics. To outline our methodology,
we proceed as follows. In Sec. \ref{sec model}, we define the model,
introduce the one-loop Luttinger--Ward (LW) functional, and derive
the corresponding self-consistent equations. Next, we describe the
main algorithm for numerically calculating the Green's functions.
These results are then used to demonstrate the existence of a spin
liquid ground state in the triangular lattice Hubbard model at the
one-loop exact level, as well as to showcase the density of states
and spectral densities for various quasiparticles. The thermodynamic
properties such as spin susceptibility, thermal conductivity, and
specific heat are discussed in Sec. \ref{sec thermodynamic}. Further,
motivated by recent experimental observation of resonant states in
cobalt atoms on single-layer 1$T$-TaSe$_{2}$ \citep{Chen2022} and
subsequent theoretical works \citep{He2022,He2023}, in Sec. \ref{sec impurity}
we use our method to analyze the Hubbard model coupled with a single
impurity. We compute the self-energy within the self-consistent first-order
Born approximation and use the Bethe-Salpeter equation to calculate
the impurity Green's function. Sec. \ref{conclusions} summarizes
the main findings and results. The details of derivations are relegated
to appendices.

\section{Luttinger--Ward functional analysis of Hubbard model using the slave
rotor construction\label{sec model}}

We study the Hubbard model on a triangular lattice, which offers a
rich framework for investigating strong electron correlations. The
Hamiltonian reads as

\begin{align}
\mathcal{H}= & \sum_{i,j,\sigma}t_{ij}c_{i,\sigma}^{\dagger}c_{j,\sigma}-\sum_{i,\sigma}\mu c_{i,\sigma}^{\dagger}c_{i,\sigma}+\frac{U}{2}\sum_{i}(n_{i,c}-1)^{2},\label{eq:Hubbard H}
\end{align}
where $c_{i,\sigma}$ ($c_{i,\sigma}^{\dagger}$) annihilates (creates)
an electron with spin $\sigma$ at site $i$, and $t_{ij}$ describes
the hopping integral between sites $i$ and $j$. $\mu$ and $U$
denote, respectively, the chemical potential and on-site Coulomb repulsion.
By employing a U(1) slave rotor decomposition \citep{Florens2002,Florens2004},
the electron operators are fractionalized into spinons and chargons
as $c_{i,\sigma}=f_{i,\sigma}X_{i}^{\dagger}$. Consequently, the
Hamiltonian is transformed to \citep{Zheng2024}

\begin{equation}
\mathcal{H}=\sum_{i,j,\sigma}t_{ij}f_{i,\sigma}^{\dagger}f_{j,\sigma}X_{j}^{\dagger}X_{i}+\mathrm{h.c.}-\sum_{i\sigma}\left(\mu+h_{i}\right)f_{i,\sigma}^{\dagger}f_{i,\sigma}+U\sum_{i}P_{i}^{\dagger}P_{i}+i\sum_{i}h_{i}P_{i}X_{i}-i\sum_{i}h_{i}X_{i}^{\dagger}P_{i}^{\dagger}+\sum_{i}\lambda_{i}\left(X_{i}^{\dagger}X_{i}-1\right)+\sum_{i}h_{i}.\label{eq:Hubbard H slave rotor}
\end{equation}
In this formulation, $f_{i,\sigma}$ ($f_{i,\sigma}^{\dagger}$) is
annihilation (creation) spinon operator. $X_{i}$ is the chargon operator
as canonical coordinate and $P_{i}$ is the momentum of scalar field
of chargon. They satisfy the commutation relations $[X_{i},P_{j}]=[X_{i}^{\dagger},P_{j}^{\dagger}]=i\delta_{ij}$.
$\lambda_{i}$ and $h_{i}$ are the Lagrange multipliers ensuring
that the constraints $X_{i}^{\dagger}X_{i}=1$ and $L_{X,i}=i(X_{i}P_{i}-X_{i}^{\dagger}P_{i}^{\dagger})=\sum_{\sigma}f_{i,\sigma}^{\dagger}f_{i,\sigma}-1$
hold.

\begin{figure}
\begin{centering}
\includegraphics[width=0.5\linewidth]{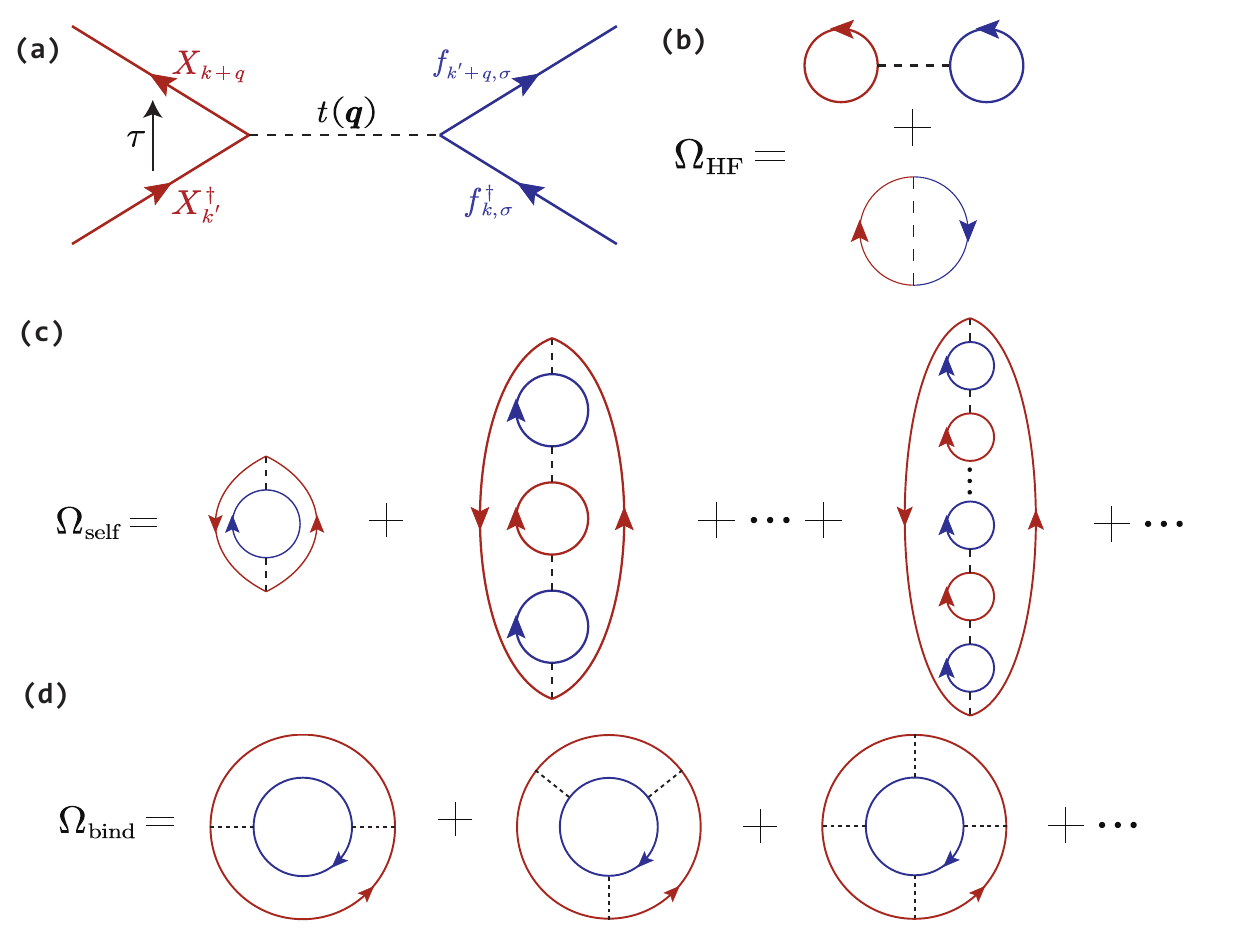} 
\par\end{centering}
\caption{One loop Luttinger--Ward functional of spinons and chargons. (a)
Spinon-chargon kinetic interaction in momentum-space $t(q)f_{k+q,\sigma}^{\dagger}X_{k^{\prime}+q}^{\dagger}f_{k^{\prime},\sigma}X_{k}$.
Red and blue lines represent the Green's functions of spinons and
chargons, respectively. (b) Hartree-Fock diagrams, which are equivalent
to the conventional mean-field theory. (c) Self-RPA polarization processes
of spinons and chargons known as the self-interaction diagrams. (d)
Polarization processes where spinons and chargons merge into electrons
known as the binding-interaction diagrams.}
\label{LW diagrams} 
\end{figure}

The above decomposition maps the original strong-coupling electron
model into a weakly coupled one described by both spinons and chargons,
where the kinetic energy term of electrons has been transformed into
a two-body term

\begin{equation}
T_{kinetic}=\sum_{ij\sigma}t_{ij}f_{i,\sigma}^{\dagger}f_{j,\sigma}X_{j}^{\dagger}X_{i}=\sum_{\boldsymbol{k},\boldsymbol{k}',\boldsymbol{q},\sigma}t(\boldsymbol{q})f_{k+q,\sigma}^{\dagger}X_{k^{\prime}+q}^{\dagger}f_{k^{\prime},\sigma}X_{k},
\end{equation}
where $t(\boldsymbol{q})=t\gamma(\boldsymbol{q})$ with $\gamma(\boldsymbol{q})=2(2\cos\frac{1}{2}q_{x}a\cos\frac{\sqrt{3}}{2}q_{y}a+\cos q_{x}a)$
as the triangular lattice form factor. We call $T_{kinetic}$ the
\textit{kinetic interaction}. To study the correlation properties
of this system, we introduce a closed set of Feynman diagrams for
two-particle processes, which define the Luttinger--Ward functional.
The kinetic interaction is diagramatically shown in Fig. \ref{LW diagrams}
(a). The Luttinger--Ward functional of spinons and chargons is written
as

\begin{equation}
\Omega_{\mathrm{LW}}=\Omega_{\mathrm{HF}}+\Omega_{\mathrm{self}}+\Omega_{\mathrm{bind}}.\label{eq:LW functional}
\end{equation}

\begin{figure}
\begin{centering}
\includegraphics[width=0.45\linewidth]{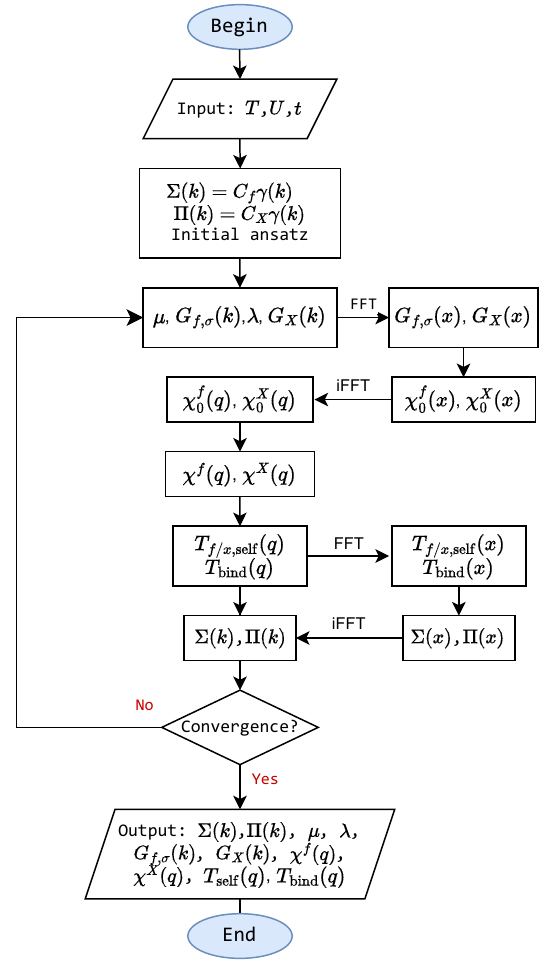} 
\par\end{centering}
\caption{Algorithm for self-consistent iterations of Green\textquoteright s
functions and self-energies of spinon and chargon. We initially set
self-nergies ansatz $\Sigma(k)=C_{f}\gamma(k)$ and $\Pi(k)=-C_{X}\gamma(k)$
which are proportional to the lattice form factor. To simplify the
one-loop susceptbility and self-energy calculations, we follow the
widely used fast Fourier transform (FFT) trick \citep{Sumita2023,Witt2021,Dee2019}
to avoid the Matsubara frequency summation in the convolution process.}
\label{flowchart} 
\end{figure}

Each term is described as follows. The quantity $\Omega_{\mathrm{HF}}$
represents the self-consistent Hartree-Fock approximation -- commonly
known as the standard mean-field theory -- and is illustrated by
the diagrams in Fig. \ref{LW diagrams}(b). The second term, $\Omega_{\mathrm{self}}$,
represents the polarization of spinons and chargons within the random
phase approximation (RPA) and is known as the self-interaction diagram.
The corresponding diagrams are shown in Fig. \ref{LW diagrams} (c).
The last term, $\Omega_{\mathrm{bind}}$, encapsulates the polarization
processes where spinons and chargons merge into electrons -- a mechanism
represented by the binding interaction diagrams shown in Fig. \ref{LW diagrams}
(d). The sum of these Feynman diagrams forms the one-loop LW functional
$\Omega_{\mathrm{LW}}$. Its relationship with the self-energy of
spinons and chargons -- as well as with the effective kinetic interaction
-- is given by the following functional derivatives:

\begin{align}
\Sigma(k)=\frac{\delta\Omega_{\mathrm{LW}}}{\delta G_{f}(k)},~~~~\Pi(k)=\frac{\delta\Omega_{\mathrm{LW}}}{\delta G_{X}(k)}\\
T_{f/X,\mathrm{self}}(k)=\frac{\delta^{2}\Omega_{\mathrm{self}}}{\delta G_{X/f}(k)\delta G_{X/f}(k)}\\
T_{\mathrm{bind}}(k)=\frac{\delta^{2}(\Omega_{\mathrm{bind}}+\Omega_{\mathrm{HF}})}{\delta G_{f}(k)\delta G_{X}(k)},\label{eq:selfenergy, interaction}
\end{align}
where $k=(\boldsymbol{k},i\omega_{n})$ is $2+1$ dimensional Matsubara
frequency-momentum vector. The fermionic and bosonic frequencies shall
be understood as $\omega_{n}=(2n+1)\pi/\beta$ and $\nu_{n}=2n\pi/\beta$,
respectively. Based on these relationships, we derive a series of
self-consistent equations that yield the Green's functions of spinons
and chargons. An overview of our computational procedure is provided
in the flowchart in Fig. \ref{flowchart}.

The Green's functions are expressed as follows:

\begin{align}
 & G_{f}^{-1}(k)=G_{f,0}^{-1}(k)-\Sigma(k),\label{eq:GFf}\\
 & G_{X}^{-1}(k)=G_{X,0}^{-1}(k)-\Pi(k),\label{eq:GF}
\end{align}
with the bare Green's functions defined as $G_{f,0}(k)=(i\omega_{n}+\mu)^{-1}$
and $G_{X,0}(k)=-(\nu_{n}^{2}+\lambda)^{-1}$. Following the steps
in the flowchart in Fig. \ref{flowchart}, the susceptibilities are
expessed as

\begin{equation}
\chi^{f/X}(q)=\frac{\chi_{0}^{f/X}(q)}{1-T_{f/X,\mathrm{self},0}(q)\chi_{0}^{f/X}(q)},\label{eq:susceptibility}
\end{equation}
where $T_{f/X,\mathrm{self},0}(q)=-t(q)\chi_{0}^{X/f}(q)t(q)$ is
defined as bare kinetic interaction and the bare suseptbilities. They
are given by $\chi_{0}^{f/X}(q)=\mp\sum_{k,\sigma}G_{f/X}(k)G_{f/X}(k+q)$.
The expressions for the effective kinetic interactions $T$ are given
by: 
\begin{align}
 & T_{f/X,\mathrm{self}}(q)=\frac{T_{f/X,\mathrm{self},0}(q)}{1-T_{f/X,\mathrm{self},0}(q)\chi_{0}^{f/X}(q)},\\
 & T_{\mathrm{bind}}(q)=\frac{t(q)}{1-t(q)G_{c,0}(q)}.\label{eq:Teff}
\end{align}

The first term in $T_{\mathrm{bind}}$ represents the mean-field contribution.
$G_{c,0}(q)=-\sum_{k}G_{f}(k)G_{X}(k+q)$ is bare electron Green's
function without corrections above second order. Additionally, the
self-energies of spinon and chargon, denoted by $\Sigma(k)$ and $\Pi(k)$,
are calculated as follows:

\begin{align}
 & \Sigma(k)=-\frac{1}{\beta N}\sum_{q}\left[T_{f,\mathrm{self}}(q)G_{f}(k+q)+T_{\mathrm{bind}}(q)G_{X}(k+q)\right],\\
 & \Pi(k)=-\frac{1}{\beta N}\sum_{q}\left[T_{X,\mathrm{self}}(q)G_{f}(k+q)+T_{\mathrm{bind}}(q)G_{f}(k+q)\right].\label{eq:selfenergy}
\end{align}

In addition to the self-consistent equations outlined above, the following
expressions enforce the slave-rotor constraints \citep{He2022,Florens2002,Florens2004}:

\begin{align}
 & 1=-\frac{1}{\beta N}\sum_{k}G_{X}(k)e^{i\nu_{n}0^{+}},\label{eq:SCE QSL lambda}\\
 & 0=-\frac{1}{2U\beta N}\sum_{k}i\nu_{n}G_{X}(k)\left[e^{i\nu_{n}0^{+}}+e^{-i\nu_{n}0^{+}}\right]+\frac{h}{U}+\frac{1}{\beta N}\sum_{k}G_{f,\sigma}(k)e^{i\omega_{n}0^{+}}-\frac{1}{2}.\label{eq:SCE QSL h}
\end{align}
Since we are interested in the spin liquid phase, the chemical potential
imposing the half-filling case is set by 
\begin{align}
1=\frac{1}{\beta N}\sum_{k,\sigma}G_{f,\sigma}(k)e^{i\omega_{n}0^{+}}.\label{eq:SCE QSL mu}
\end{align}

The Eq. \eqref{eq:SCE QSL h} has a solution for $h=0$, thus eliminating
the need for further calculation. We solve the self-consistent Eqs.
\eqref{eq:GF}-\eqref{eq:SCE QSL mu} to obtain the spectral functions
of spinons and chargons from the corresponding Green's functions.
For numerical calculations we set parameters $t=0.091~\mathrm{eV}$
and $U=0.775~\mathrm{eV}$ \citep{He2022,Zheng2024} under Brillouin
zone mesh $N=100\times100$. For transition from imaginary to real
frequencies, we employ intermediate representation (IR) basis \citep{Otsuki2020,Shinaoka2022,Shinaoka2017,Li2020}
to store and calculate Matsubara Green's function, and perform Nevanlinna
analytical continuation (NAC) \citep{Nogaki2023a,Nogaki2023,Fei2021}
to achieve real frequency Green's function numerically.

\begin{figure}
\begin{centering}
\includegraphics[width=0.5\linewidth]{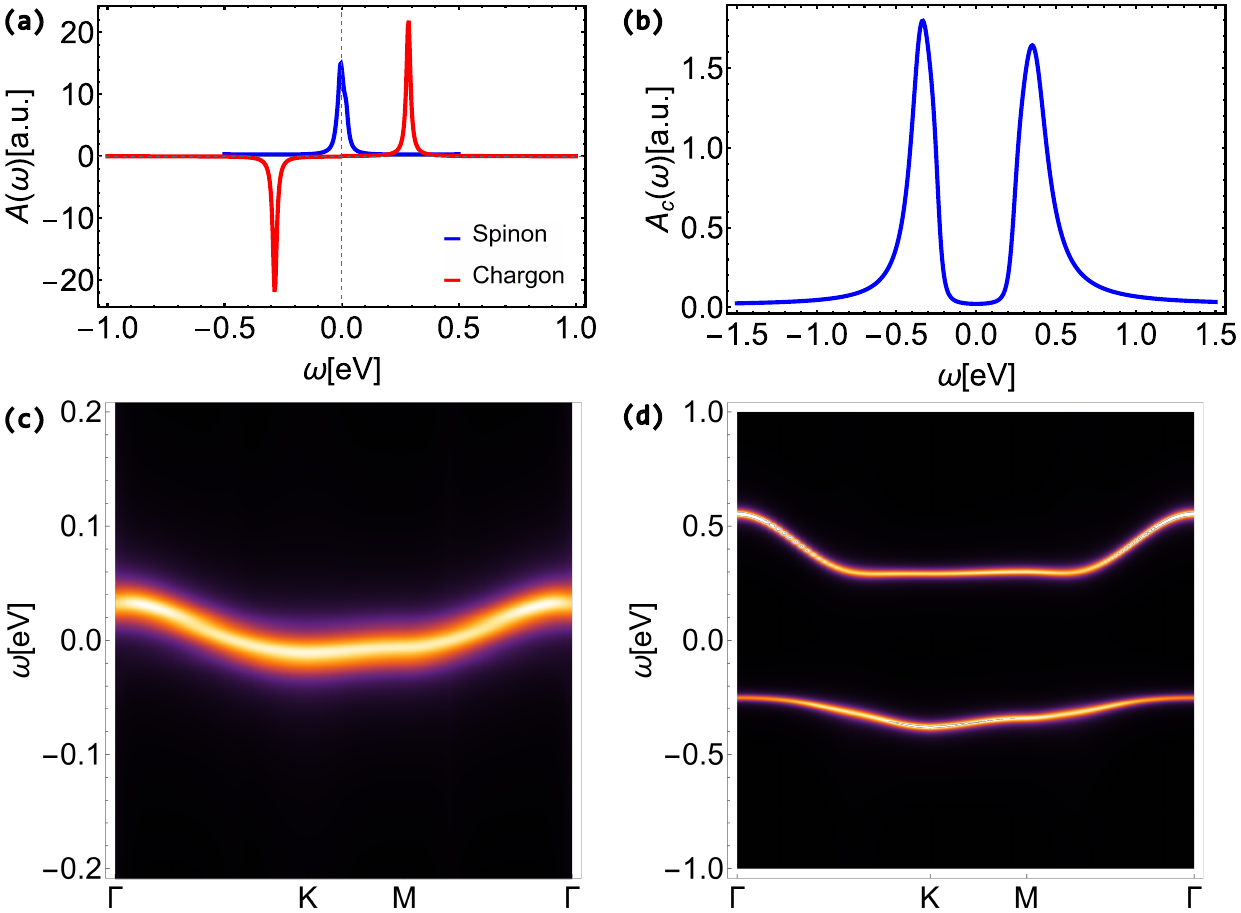} 
\par\end{centering}
\caption{Density of states of (a) spinon and chargon and (b) spinon-chargon
bound state, namely the physical electron. Spectral density representing
the excitations of (c) spinon and (d) the physical electron. The temperature
here is set $T=0.0001~\mathrm{K}$.}
\label{LW QSL GF} 
\end{figure}

Not only does the binding kinetic interaction between spinons and
chargons, $T_{\mathrm{bind}}$, contribute to the aforementioned self-energy,
but it also merges them into a bound state -- that is, the physical
electron. This process is described by ladder diagrams in the Bethe-Salpeter
equation (BSE). However, employing the BSE with non-zero center-of-mass
momentum in the IR basis poses significant challenges. Fortunately,
since our kinetic interactions originate from electron hopping, their
form simplifies the BSE on the lattice. Hence, it is more convenient
to consider the BSE on a real-space lattice with imaginary time:

\begin{equation}
G_{c}\left(x_{1}^{\prime\prime\prime},x_{2}^{\prime\prime\prime};x_{1},x_{2}\right)=G_{c,0}\left(x_{1}^{\prime\prime\prime},x_{2}^{\prime\prime\prime};x_{1},x_{2}\right)-\sum_{\begin{array}{c}
x_{1}^{\prime},x_{2}^{\prime}\\
x_{1}^{\prime\prime},x_{2}^{\prime\prime}
\end{array}}G_{c,0}\left(x_{1}^{\prime\prime\prime},x_{2}^{\prime\prime\prime};x_{1}^{\prime\prime},x_{2}^{\prime\prime}\right)K^{*}\left(x_{1}^{\prime\prime},x_{2}^{\prime\prime};x_{1}^{\prime},x_{2}^{\prime}\right)G_{c}\left(x_{1}^{\prime},x_{2}^{\prime};x_{1},x_{2}\right).\label{eq:QSL BSE 1}
\end{equation}
Here, $G_{c}\left(x_{1}^{\prime},x_{2}^{\prime};x_{1},x_{2}\right)$
is the general two-body Green's function for the electron bound state,
with two incoming particles with spacetime coordinates $x_{1},x_{2}$
and outgoing coordinates $x_{1}^{\prime},x_{2}^{\prime}$. Since in
the original slave rotor theory the electron operator is defined as
$c_{i,\sigma}=f_{i,\sigma}X_{i}^{\dagger}$ , the electron bound state
is formed solely by combining the spinon and chargon at the same lattice
site. Consequently, the BSE reduces to

\begin{equation}
G_{c}\left(x^{\prime\prime\prime},x\right)=G_{c,0}\left(x^{\prime\prime\prime},x\right)+\sum_{x^{\prime},x^{\prime\prime}}G_{c,0}\left(x^{\prime\prime\prime},x^{\prime\prime}\right)t(x^{\prime\prime},x^{\prime})G_{c}\left(x^{\prime},x\right).\label{eq:QSL BSE 2}
\end{equation}
In this equation, we substitute the first-order interaction kernel
$K^{*}\left(x_{1}^{\prime\prime},x_{2}^{\prime\prime};x_{1}^{\prime},x_{2}^{\prime}\right)=-t(\boldsymbol{r}_{2}^{\prime},\boldsymbol{r}_{1}^{\prime})\delta(\boldsymbol{r}_{1},\boldsymbol{r}_{2})\delta(\boldsymbol{r}_{1}^{\prime},\boldsymbol{r}_{2}^{\prime})$
into the BSE. As a result, the electron Green's function can be readily
obtained as:

\begin{equation}
G_{c}(x^{\prime\prime},x)=\sum_{x^{\prime}}\left[1-G_{c,0}t\right]^{-1}(x^{\prime\prime},x^{\prime})G_{c,0}(x^{\prime},x).\label{eq:QSL BSE 3}
\end{equation}
Using the Green's functions $G_{f}$, $G_{X}$ and $G_{c}$, we show
in Fig. \ref{LW QSL GF} the density of states and spectral densities
by computing $A(\boldsymbol{k},\omega)=-\frac{1}{\pi}\mathrm{Im}\left[G(k,i\omega_{n}\rightarrow\omega+i0^{+})\right]$.
Fig. \ref{LW QSL GF}(a) shows the density of states for spinons and
chargons separately. In Figure \ref{LW QSL GF}(b), the electron density
of states reveals clear peaks corresponding to the Mott-Hubbard energy
bands. Figure \ref{LW QSL GF}(c) illustrates the spinon spectrum,
indicating the presence of a spinon Fermi surface. As expected, the
electron spectrum -- the bound state formed by spinons and chargons
-- exhibits the Mott-Hubbard energy bands, as seen in Figure \ref{LW QSL GF}(d).
Notably, these results are obtained without invoking gauge fields,
yet the spectrum closely resembles those derived using conventional
mean-field theory supplemented by gauge field fluctuation methods
\citep{He2023a,Zheng2024}.

\begin{table}
\begin{centering}
\begin{tabular}{|c|c|}
\hline 
{\footnotesize{}{}}\textbf{\footnotesize{}Methods}  & {\footnotesize{}{}}\textbf{\footnotesize{}Critical interaction}{\footnotesize{}
$U_{c}/t$}\tabularnewline
\hline 
\hline 
{\footnotesize{}{}Slave-rotor MFT}  & {\footnotesize{}{}5.16}\tabularnewline
\hline 
{\footnotesize{}{}SRLW Hartree-Fock {[}this work{]}}  & {\footnotesize{}{}5.145 - 5.15}\tabularnewline
\hline 
{\footnotesize{}{}SRLW one-loop {[}this work{]}}  & {\footnotesize{}{}6.7 - 6.75}\tabularnewline
\hline 
{\footnotesize{}{}DMRG \citep{Shirakawa2017}}  & {\footnotesize{}{}7.55 - 8.05}\tabularnewline
\hline 
{\footnotesize{}{}DMRG \citep{Szasz2020}}  & {\footnotesize{}{}8.0}\tabularnewline
\hline 
{\footnotesize{}{}Exact diagonlization \citep{Clay2008}}  & {\footnotesize{}{}7.0}\tabularnewline
\hline 
{\footnotesize{}{}Exact diagonlization \citep{Kokalj2013}}  & {\footnotesize{}{}7.4 - 7.5}\tabularnewline
\hline 
{\footnotesize{}{}Variational cluster approximation \citep{Sahebsara2008}}  & {\footnotesize{}{}6.7}\tabularnewline
\hline 
{\footnotesize{}{}Variational cluster approximation \citep{Yamada2014}}  & {\footnotesize{}{}6.7 (6.3 for 12D cluster)}\tabularnewline
\hline 
\end{tabular}
\par\end{centering}
\caption{Comparison of critical interaction strengths for the Mott transition
obtained by various methods at zero temperature. The first-row entry
was computed using the conventional slave-rotor mean-field theroy.
The second and third rows report results from the method proposed
in this paper: the second-row corresponds to the self-consistent Hartree--Fock
diagram (equivalent to mean-field level; see Fig. \ref{LW diagrams}b),
while the third row corresponds to the one-loop diagram (see Fig.
\ref{LW diagrams}c-d). }
\label{Benchmark} 
\end{table}

\begin{figure}
\begin{centering}
\includegraphics[width=0.5\linewidth]{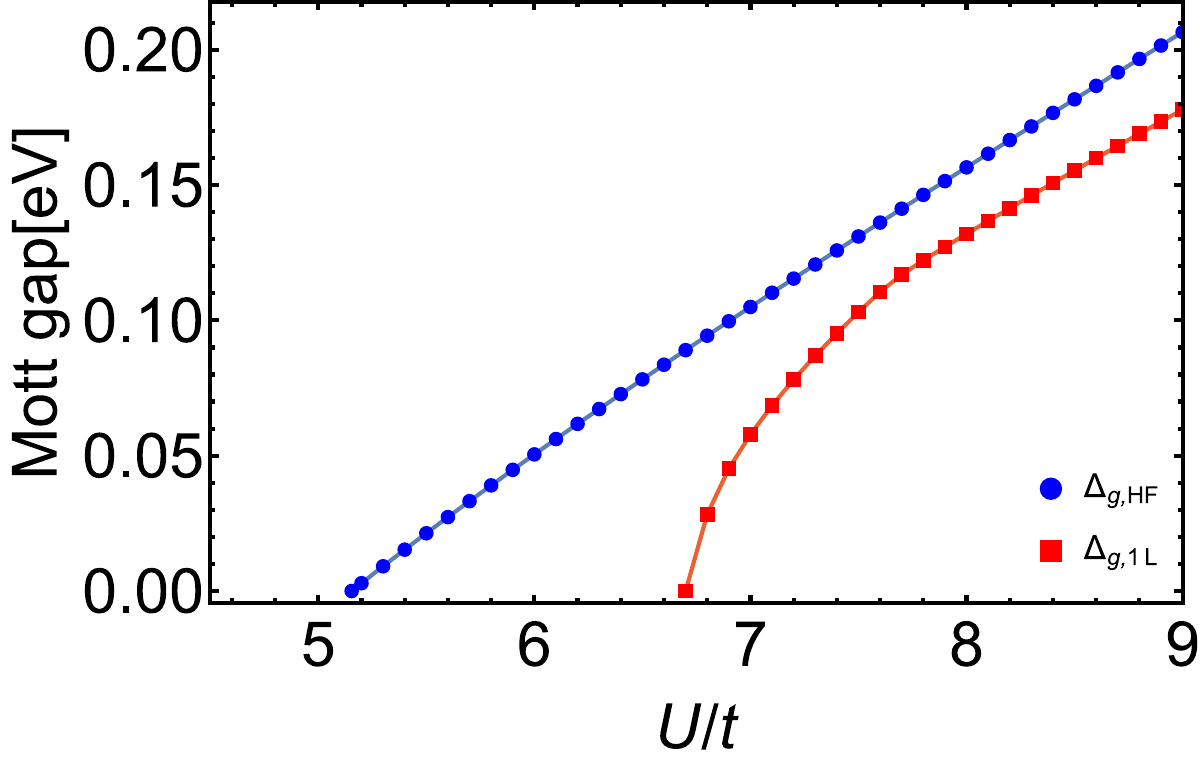} 
\par\end{centering}
\caption{Variation of the Mott gap as a function of $U/t$ evaluated using
the self-consistent Hartree-Fock calculations (blue solid circles)
and the self-consistent Luttinger--Ward one-loop calculations (red
solid squares).}
\label{Mottgap} 
\end{figure}

Before concluding this section, we compare the Mott transition in
the Hubbard model on triangular lattice -- obtained using our method
outlined above -- with results from other studies in the literature.
In Table \ref{Benchmark} we compare the onset of Mott transition
obtained by using different approaches. Our method in contrast to
the mean-field theory yields a critical value (in the third row and
see diagrams in Fig. \ref{LW diagrams}c-d)) being close to the values
obtained in other numerical approaches. When the calculations are
restricted to the Hartree-Fock level -- corresponding to the diagrams
in Fig. \ref{LW diagrams}b of the main text -- our method reproduces
the mean-field results, as expected; see the values reported in the
first and second rows of Table \ref{Benchmark}. Further, in Fig.
\ref{Mottgap} we present the variation of the Mott gap as a function
of $U/t$. Our findings clearly depart from those of mean-field theory.

\section{thermodynamic properties\label{sec thermodynamic}}

The thermodynamic characterization of U(1) quantum spin liquids is
essential for unveiling their low-energy effective field theories
and underlying quantum orders. In this section, we analyze key thermodynamic
observables -- such as spin susceptibility, thermal conductivity,
and specific heat -- to elucidate the interplay between fluctuations
in kinetic interactions and spinon correlations within these systems.
We compute the corresponding response functions using the properties
of Matsubara Green's functions and linear response theory. Investigating
the physical properties of QSLs has long been challenging, primarily
because the quasiparticle excitations -- the spinons -- are charge-neutral.
Leveraging the efficiency of the IR basis to simplify Green's function
calculations, we evaluate the following physical quantities: 
\begin{itemize}
\item Spin susceptibility of spinons: Since spinons carry the intrinsic
spin degrees of freedom of fractionalized electrons, their spin susceptibility
directly reflects the collective magnetic excitations within the spin
liquid. By analyzing this response, one can gain valuable insights
into the underlying interactions governing the spin excitations.


\item Thermal conductivity and specific heat: Determining the nature of
QSLs often hinges on a detailed comparison of their thermodynamic
and thermal transport properties. These properties are particularly
contentious for U(1) QSLs with a spinon Fermi surface. Here, we incorporate
the temperature dependence of the self-energies to calculate the spinon
thermal conductivity and specific heat. These calculations aim to
provide insights into the interplay between kinetic interaction fluctuations
and spinon correlations, further illuminating the low-energy behavior
of these systems. 
\end{itemize}

\subsection{Spin susceptibility}

The spin structure factor $S^{\alpha\beta}(q)$ is related to the
spin relaxation rate as $\tau_{1}^{-1}\propto\sum_{\boldsymbol{q}}S(q)$.
Furthermore, the $S^{\alpha\beta}(q)$ and the spin susceptibility
tensor $\chi_{\mathrm{spin}}^{\alpha\beta}$ is related by

\begin{equation}
S^{\alpha\beta}(q)=\frac{-2\mathrm{Im}\chi_{\mathrm{spin}}^{\alpha\beta}(q)}{1-e^{-\beta\omega}},\label{eq:spin structurefactor}
\end{equation}
where $\chi_{\mathrm{spin}}^{\alpha\beta}(k)=-\frac{1}{N}\sum_{i}\int_{0}^{\beta}d\tau e^{i\nu_{n}\tau-i\boldsymbol{k}\cdot\boldsymbol{r}_{i}}\left\langle \mathrm{S}^{\alpha}(\boldsymbol{r}_{i},\tau)\mathrm{S}^{\beta}(0,0)\right\rangle $
and the spin operator is defined as $\mathrm{S}^{\alpha}(\boldsymbol{r}_{i})=\frac{1}{2}f_{i,\varsigma}^{\dagger}\sigma_{\varsigma\varsigma^{\prime}}^{\alpha}f_{i,\varsigma^{\prime}}$
\citep{Scheurer2018}.

\begin{figure}
\begin{centering}
\includegraphics[width=0.5\linewidth]{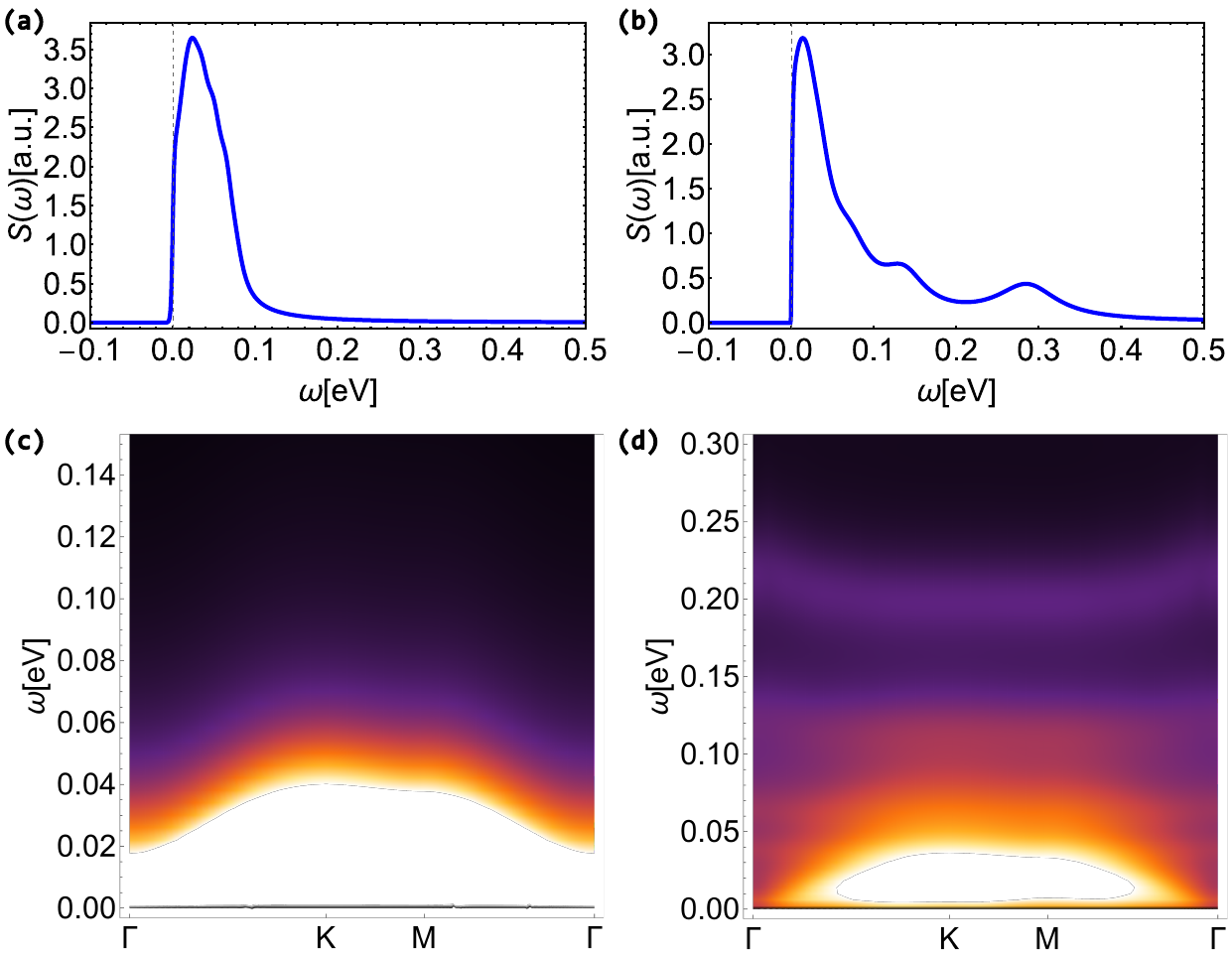} 
\par\end{centering}
\caption{Spin structure factor $S(\omega)$ calculated by considering (a) only
the Hartree-Fock diagrams and (b) incorporates the kinetic effective
self-interactions on top of the Hartree-Fock result. The spectral
density of the spin susceptibility along high-symmetry paths in momentum
space using only (c) Hartree-Fock diagrams (d) including the self-interaction
corrections. The temperature here is $T=0.001~\mathrm{K}$.}
\label{spin susceptibility} 
\end{figure}

The system is spin-degenerate, so we focus solely on the longitudinal
component of the susceptibility. This allows us to employ Eq. \eqref{eq:susceptibility}
to compute the spin susceptibility and the corresponding spin structure
factor using the one-loop exact spinon Green's function. Figure \ref{spin susceptibility}
shows the spin susceptibility, $S(\omega)$, alongside the spin spectral
density, $-\frac{1}{\pi}\mathrm{Im}\chi_{\mathrm{spin}}^{\alpha\beta}(\boldsymbol{q},\omega)$.
At the mean-field level, as depicted in panels (a) and (c), only a
single spectrum is observed at low energies. However, once the kinetic
interaction is taken into account, additional features appear at higher
energies manifested as broad peaks in $S(\omega)$ (Fig. \ref{spin susceptibility}(b))
and as nearly dispersionless collective excitations in the spin spectral
density (Fig. \ref{spin susceptibility}(d)).


\begin{figure*}
\begin{centering}
\includegraphics[width=1\linewidth]{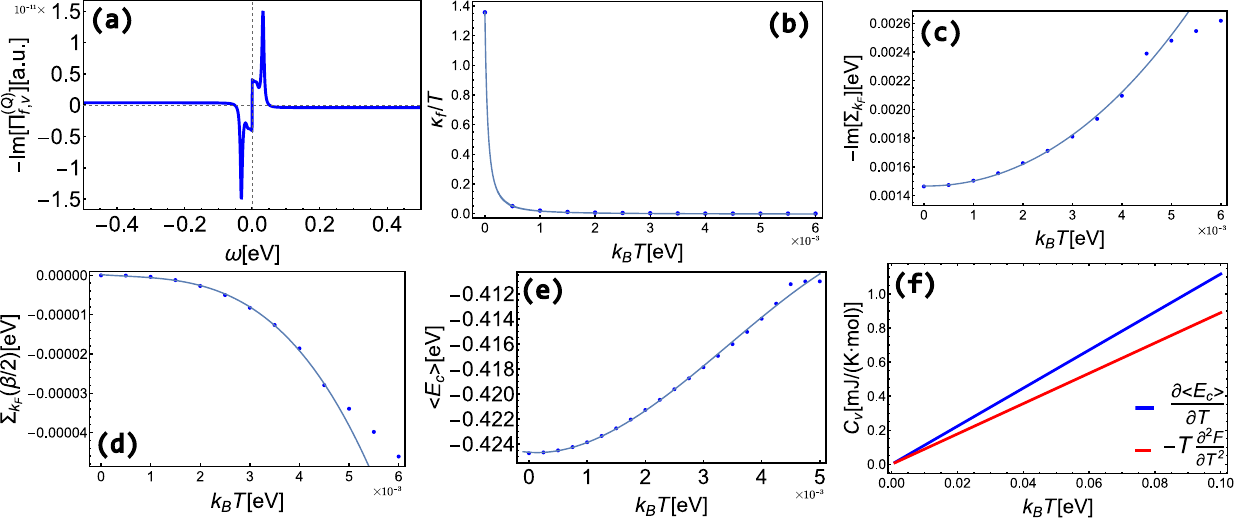} 
\par\end{centering}
\caption{(a) Imaginary part of the the one-loop corrected spinon heat current-current
correlation function $-\frac{1}{\pi}\mathrm{Im}[\Pi_{f,v}^{(Q)}(\boldsymbol{q}\rightarrow0,\omega)]$.
(b) Temperature variation of spinon thermal conductivity $\kappa/T$
calculated by analytical continuation of $\Pi_{f,v}^{(Q)}(i\nu_{1})$.
Here the blue dots correspond to numerical data fitted by $a+b_{1}/T+b_{2}/T^{2}$.
(c) Imaginary part of the retarded spinon self-energy $\Sigma(\boldsymbol{k}=\boldsymbol{k}_{F},\omega=0)$
as a function of $T$. (d) Imaginary time spinon self-energy $\Sigma(\boldsymbol{k}=\boldsymbol{k}_{F},\tau=\beta/2)$,
valued at Fermi surface and $\tau=\beta/2$. (e) The electron internal
energy computed by imaginary time derivative of electron Green's function,
resulting in $T^{2}$ behavior in very low temperature regime. (f)
The temperature dependence of the specific heat, computed using the
electronic internal energy (solid blue line) and the system's average
free energy (solid red line).}
\label{thermal conductivity} 
\end{figure*}

\subsection{Thermal conductivity}

In the absence of mobile charge carriers, thermal conductivity is
predominantly governed by spinon contributions, thereby revealing
the gapless nature of the spinon Fermi surface. The thermal conductivity
is defined by \citep{Mahan2000,Werman2018,Fabrizio2023}:

\begin{equation}
\kappa_{f}(T)=-\frac{1}{T}\lim_{\omega\rightarrow0}\frac{\mathrm{Im}[\Pi_{f}^{(Q)}(\boldsymbol{q}=0,\omega)]}{\omega},\label{eq:thermal conductivity}
\end{equation}
where $\Pi_{f}^{(Q)}(\boldsymbol{q},\omega)$ is the $xx$-component
of spinon heat current-current correlation function. To account for
the effects of interactions, the heat flow correlation function is
separated into the free and interaction parts: $\Pi_{f}^{(Q)}=\Pi_{f,0}^{(Q)}+\Pi_{f,v}^{(Q)}$.
The interaction part $\Pi_{f,v}^{(Q)}$ is computed as \citep{Khatibi2020}:
\begin{equation}
\Pi_{f,v}^{(Q)}(q)=-\Pi_{f,lc}^{(Q)}(q)T_{\mathrm{self}}(q)\Pi_{f,rc}^{(Q)}(q),
\end{equation}
where the left and right vertices are:

\begin{equation}
\Pi_{f,lc}^{(Q)}(q)=\Pi_{f,rc}^{(Q)}(q)=-\frac{1}{\beta N}\sum_{k}\Gamma_{x}^{(Q)}(\boldsymbol{k},\boldsymbol{k}+\boldsymbol{q})G_{c}(k)G_{c}(k+q)
\end{equation}
with bare vertex $\Gamma_{x}^{(Q)}(\boldsymbol{k},\boldsymbol{k}+\boldsymbol{q})=\frac{1}{2}\left(\Xi_{f}(\boldsymbol{k}+\boldsymbol{q})v_{f,x}(\boldsymbol{k})+\Xi_{f}(\boldsymbol{k})v_{f,x}(\boldsymbol{k}+\boldsymbol{q})\right)$,
where $\Xi_{f}(\boldsymbol{k})=\Sigma(\boldsymbol{k})-\mu$ is the
relative energy of spinon and $v_{f,x}(\boldsymbol{k})=\left.\frac{\partial\Xi_{f}}{\partial k_{x}}\right|_{\boldsymbol{k}}$
denotes the velocity of spinon. The details are given in Appendix
\ref{Appendix Current operator}.

As shown in Fig. \ref{thermal conductivity}(a), the spectral density
of the spinon thermal current-current correlation function as a function
of frequency clearly reveals both a Drude peak and side peaks induced
by $T_{\mathrm{self}}$, which is reminiscent of a typical Fermi liquid
\citep{Gunnarsson2010,Mei2022,Huang2023}. Meanwhile, in order to
investigate the thermal transport properties of the spin liquid, it
is crucial to examine the temperature dependence of the spinon thermal
conductivity---this quantity often reflects the type of spin liquid
and the characteristics of its low-energy excitations. Fig. \ref{thermal conductivity}(b)
presents the variation of the spinon thermal conductivity, with increasing
temperature. The fitting of the numerical data indicates that the
spinon thermal conductivity exhibits an inverse square dependence
on temperature. This result is in contrast to earlier theoretical
predictions that the thermal conductivity of a clean U(1) spin liquid
with a Fermi surface should scale as $T^{1/3}$ \citep{Werman2018,Nave2007}.

The above conclusion can be further coroborated by examining the temperature
dependence of the spinon self-energy at low energies. As depicted
in Fig. \ref{thermal conductivity} (c), a clear cubic relationship
emerges for the imaginary part of real-time spinon self-energy, which
located near Fermi surface, $\boldsymbol{k}=\boldsymbol{k}_{F}$.
Here, we are concern about the contribution of effective kinetic interaction
scattering to the spinon thermal transport. The scattering rate is
directly connected to the imaginary part of the self-energy via $\Gamma_{f}=-2\mathrm{Im}[\Sigma]\propto T^{3}$
which implies $\kappa_{f}\propto C_{V}v_{f}^{2}\Gamma_{f}^{-1}\propto T^{-2}$.
Hence, the spinon thermal conductivity indeed scales inversely with
the square of the temperature. On the other hand, to avoid potential
numerical errors arising from Nevanlinna analytical continuation in
the calculation of the spinon self-energy, we have simultaneously
computed the imaginary-time self-energy $\Sigma(\tau)$ at $\tau=\beta/2$,
thereby providing a qualitative estimate of the magnitude of the retarded
self-energy $\mathrm{Im}[\Sigma(0+i0^{+})]$ at zero frequency \citep{Huang2019},
as shown in Fig. \ref{thermal conductivity}(d). It is ultimately
found that this quantity likewise exhibits a $T^{3}$ temperature
dependence, consistent with the results obtained via NAC.

\subsection{Specific heat}

One of the key physical observables for probing and characterizing
spin liquids is the temperature-dependent behavior of internal energy
and specific heat. Using the electron Green's function obtained in
Eq. \eqref{eq:QSL BSE 3}, the average electron internal energy is
calculated as $\left\langle E_{c}\right\rangle =\partial_{\tau}G_{c}(\boldsymbol{r},\tau\rightarrow0^{+})$
\citep{Fetter2003}, which increases quadratically with temperature
at low $T$, as illustrated in Fig. \ref{thermal conductivity}(e).
According to the definition of the electronic specific heat, $C_{V}=\left.\frac{\partial\left\langle E_{c}\right\rangle }{\partial T}\right|_{V}$,
we obtain the linear-$T$ increasing specific heat result from fitted
datas of average electron energy, which is shown in Fig. \ref{thermal conductivity}(f).
For comparison, we have also computed the specific heat by evaluating
the system's thermodynamic potential and free energy within the definition
by LW functional. From the definition $C_{v}=-T\frac{\partial^{2}F}{\partial T^{2}}$
, the outcome agrees closely with the result obtained from derivative
the average internal energy and likewise reveals a specific heat that
grows $T$-linearly. While both methods yield the correct linear behavior,
the discrepancy arises from insufficient knowledge of the long-tail
characteristics of the Matsubara functions. Detailed derivations are
provided in the Appendix \ref{Appendix Thermodynamic potential}.

To elaborate further on our findings and those obtained via conventional
methods, we present the following comments. 
\begin{enumerate}
\item In conventional approaches, a mean-field saddle-point solution is
adopted with gauge fluctuations treated at the quadratic level. In
this framework, the inverse of the transverse gauge-field Green's
function is expressed as $D_{ij}^{-1}(q)=\Pi_{f,ij}+\Pi_{X,ij}$,
where $\Pi_{f,ij}$ and $\Pi_{X,ij}$ denote the $ij$ components
of the spinons and chargons (or holons) current-current correlation
function, respectively \citep{Lee1992,Nagaosa1990,Nave2007}. In the
long-wavelength limit, the transverse gauge-field propagator vanishes
as $q^{3}$. Consequently, its interaction with the spinons generates
a self-energy with both real and imaginary parts scaling as $\textrm{Re}[\Sigma]\sim\textrm{Im}[\Sigma]\sim\omega^{2/3}$.
This behavior leads directly to a divergent density of states and
effective mass, a specific heat that scales as $T^{2/3}$, and a thermal
conductivity varying as $T^{1/3}$. 
\item In this work, we derive the spinon self-energy from Eqs. \eqref{eq:Teff}-\eqref{eq:selfenergy}.
The effective kinetic interaction, $T_{f,\mathrm{self}}(q)$, defined
relative to the Hartree-Fock diagrams at the mean-field level, plays
a role analogous to the conventional RPA correction in an interacting
Fermi gas. The poles in the high-frequency regime correspond to collective
spinon excitations, as illustrated in Fig. \ref{spin susceptibility}
(b) and (d). Importantly, the absence of divergence in the low-energy
density of states ensures that the spinons possess a well-defined,
stable Fermi surface. 
\item In \citep{Li2021}, it was demonstrated that a proper treatment of
the emergent gauge field restricts its low-energy excitations to contribute
no more than a $T^{2}$ term to the specific heat. This result rules
out the previously reported anomalous behaviors--namely, the $T^{2/3}$
variation in specific heat and the $T^{1/3}$ scaling of thermal conductivity.
Consequently, in the U(1) spinon Fermi surface spin liquid considered
here, both the specific heat and the thermal conductivity follow conventional
Fermi liquid behavior. Here, the self-consistent, one-loop LW numerical
calculations show that the specific heat of the U(1) spinon Fermi-surface
spin liquid grows linearly with temperature, in close agreement with
Fermi-liquid theory.


\end{enumerate}

\section{Spinon Kondo effect and the role of kinetic interaction\label{sec impurity}}

To further validate and extend the applicability of our method, in
this section we study a spin liquid phase coupled to dilute magnetic
impurities. This study is motivated by recent experimental observation
of resonant states in cobalt atoms on single-layer 1$T$-TaSe$_{2}$
\citep{Chen2022}. A slave-rotor analysis of the model supplemented
by gauge fields can account for the resonant states \citep{He2022,He2023}.
Here, we reconsider the problem and treat it using the approach stated
in precending section. The Hubbard model coupled with a magnetic impurity
reads as

\begin{equation}
\mathcal{H}_{\mathrm{hybrid}}=\sum_{i,j,\sigma}t_{ij}c_{i,\sigma}^{\dagger}c_{j,\sigma}+\sum_{\sigma}\epsilon_{d}d_{\sigma}^{\dagger}d_{\sigma}+V\sum_{\sigma}c_{0,\sigma}^{\dagger}d_{\sigma}+\mathrm{h.c.}+\frac{U}{2}\sum_{i}(n_{i,c}-1)^{2}+\frac{U_{\mathrm{imp}}}{2}(n_{d}-1)^{2},\label{eq:H hybrid}
\end{equation}
In this equation, parameter $V$ is the local impurity-host electron
hybridize and $U_{\mathrm{imp}}$ is local impurity interaction. Similarly,
by replacing the electron operators with slave particles as $c_{i,\sigma}=f_{i,\sigma}X_{i}^{\dagger}$,
$d_{i,\sigma}=a_{i,\sigma}Y_{i}^{\dagger}$, in this representation
we obtain the impurity hybridized Hamiltonian $\mathcal{H}_{\mathrm{hybrid}}=\mathcal{H}+\mathcal{H}'$,
where $\mathcal{H}$ is the Hamiltonian of spin liquid in Eq. \eqref{eq:Hubbard H slave rotor}
and $H'$ describes the impurity and hybridization with the spin liquid:

\begin{equation}
\mathcal{H}'=\left(\epsilon_{d}-h_{2}\right)a_{\sigma}^{\dagger}a_{\sigma}+U_{\mathrm{imp}}Q^{\dagger}Q+\lambda_{2}\left(Y^{\dagger}Y-1\right)+h_{2}+V\sum_{\sigma}f_{\sigma}^{\dagger}a_{\sigma}Y^{\dagger}X+h.c.
\end{equation}
The operator $Q$ is canonical momentum of $Y$ which satisfies $[Q,Y]=i$.
Additionally, the parameters $\lambda_{2}$ and $h_{2}$ are Lagrange
multiplies, constraining the norm of impurity chargon $Y$ and impurity
angular momentum.

In terms of coupled system, it is straightforward to define the joined
field basis: $\psi=(a,f)^{T}$ and $\phi=(Y,X)^{T}$. Then the self-energies
of coupled system take forms of symmetric off-diagonal matrices: $\Sigma_{\mathrm{hybrid}}=-iw\sigma_{y}$,
$\Pi_{\mathrm{hybrid}}=iu\sigma_{y}$ where $\sigma_{y}$ denotes
the $y$-Pauli matrix. Finally, the self-consistent equations within
the first-order Born approximation are derived as (the details of
derivation are given in Appendix \ref{Appendix Impurity}):

\begin{align}
u= & -\frac{2V}{\beta}\sum_{n}G(a,f^{\dagger},i\omega_{n},\sigma),\label{eq: SCE u}\\
w= & -\frac{V}{\beta}\sum_{n}G(X,Y^{\dagger},i\nu_{n}),\label{eq: SCE w}\\
1= & -\frac{1}{\beta}\sum_{n}G(Y,Y^{\dagger},i\nu_{n})e^{i\nu_{n}0^{+}},\label{eq: SCE lambda2}\\
0= & -\frac{1}{2U\beta}\sum_{n}i\nu_{n}G(Y,Y^{\dagger},i\nu_{n})\left[e^{i\nu_{n}0^{+}}+e^{-i\nu_{n}0^{+}}\right]+\frac{h_{2}}{U}+\frac{1}{\beta}\sum_{n}G(a,a^{\dagger},i\omega_{n},\sigma)e^{i\omega_{n}0^{+}}-\frac{1}{2}.\label{eq: SCE h2}
\end{align}
In equations Eqs \eqref{eq: SCE w}-\eqref{eq: SCE h2}, $u$ and
$w$ represent the impurity scattering self-energies for chargon and
spinon respectively. Additionally, Eqs \eqref{eq: SCE lambda2} and
\eqref{eq: SCE h2} correspond to the constraints on the norm and
angular momentum of the chargons, respectively.

\begin{figure}
\begin{centering}
\includegraphics[width=0.5\linewidth]{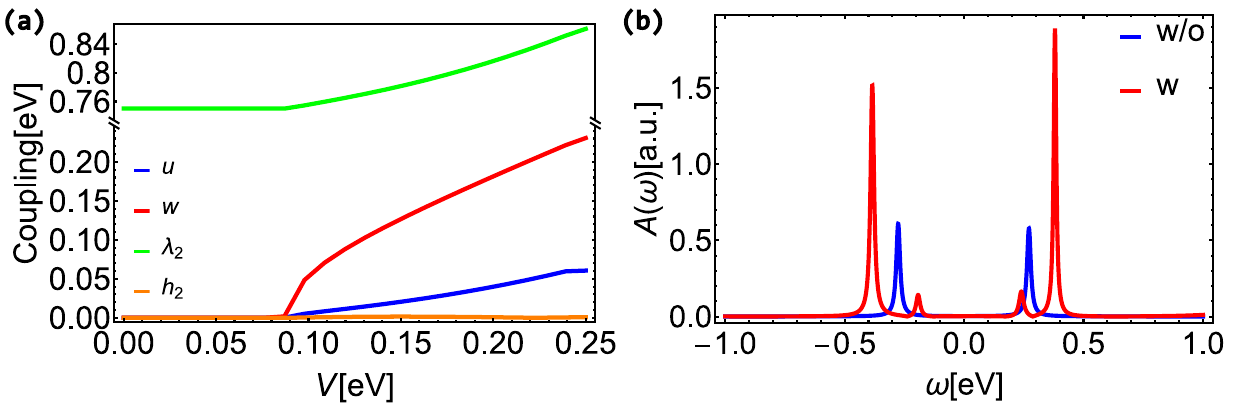} 
\par\end{centering}
\caption{(a) Variation of coupling and constraint fields in the first order
self-consistent Born approximation. Datas are shown as a function
of increasing $V$ with $U_{\mathrm{imp}}=3~\mathrm{eV}$. (b) Impurity
electron spectral functions with (red) and without (blue) the contribution
of binding interaction $T_{\mathrm{bind}}$, when $V=0.2~\mathrm{eV}$.
Close to the main Hubbards bands the resonant peaks appear due to
the bindig interaction. }
\label{impurity GF} 
\end{figure}

The variation of effective couplings $u$ and $w$ is shown in Fig.
\ref{impurity GF}(a). It is evident that as the impurity hybridization
strength $V$ increases, a critical point emerges. For values of $V$
below this threshold, the impurity decouples from the spin liquid
\citep{Chen2022,He2022}. However, beyond the critical point a hybridized
phase sets in; in this phase, the self-energies $u$ and $w$ as well
as the angular momentum Lagrange multiplier $h_{2}$ acquire nonzero
values. Moreover, the chargon amplitude Lagrange multiplier $\lambda_{2}$
grows with $V$, a consequence of the local impurity no longer being
restricted to single occupancy.

Within our approach the resonant states, as observed experimentally
in \citep{Chen2022}, are captured by computing the local Green's
function taking into account binding interaction $T_{\mathrm{bind}}$
derived from the $\Omega_{\mathrm{LW}}$. As mentioned before, the
$T_{\mathrm{bind}}$ fuses spinons and chargons into bound states
forming electrons. We calculate the impurity electron Green's function
corrected by the BSE effective interaction:

\[
G_{d}(i\omega_{n})=G_{d,0}(i\omega_{n})-[G(a,f^{\dagger})\ast G(X,Y^{\dagger})](i\omega_{n})T_{\mathrm{bind}}(i\omega_{n},0)[G(f,a^{\dagger})*G(Y,X^{\dagger})](i\omega_{n}).
\]
Here, $G_{d,0}(i\omega_{n})=-[G(a,a^{\dagger})\ast G(Y,Y^{\dagger})](i\omega_{n})=-\frac{1}{\beta}\sum_{\nu_{n}}G(a,a^{\dagger},i\omega_{n}+i\nu_{n})G(Y,Y^{\dagger},i\nu_{n})$
is the impurity electron Green's function. The symbol $*$ denotes
convolution in Matsubara-momentum space. We present the local impurity
density of states in Fig. \ref{impurity GF} (b), where we considered
the cases with and without the corrections yielded by kinetic interaction
$T_{\mathrm{bind}}$. As seen, the differences are significant. The
inclusion of kinetic interaction leads to the formation of resonant
peaks, while in for the case of no kinetic interaction only the local
Hubbard states are observed.

\section{CONCLUSIONs\label{conclusions}}

In this work, we have developed a unified framework combining the
Luttinger--Ward functional approach with the slave rotor formalism
to investigate the U(1) quantum spin liquid (QSL) phase in the triangular
lattice Hubbard model. By systematically constructing the one-loop
LW functional for spinons and chargons, we derived self-consistent
equations for Green's functions while incorporating gauge field fluctuations
beyond conventional mean-field approximations. Our results demonstrate
the emergence of a spin liquid ground state characterized by a spinon
Fermi surface and gapped chargon excitations, consistent with spin-charge
separation in strongly correlated systems.

The main findings are summarized as follows: (1) at the level of self-consistent
one-loop truncation, the triangular lattice Hubbard model with slave
rotor construction admits a U(1) spin liquid with a spinon Fermi surface
ground state. (2) in this approach the effective kinetic interaction
$T_{\text{bind}}$ (equivalent to emergent U(1) photons) mediates
spinon-chargon binding processes, enabling the reconstruction of electronic
spectral functions through Bethe-Salpeter equations. 
(3) the spinon-dominated thermal conductivity exhibits a linear behavior
in $T^{-2}$, which is in contrast to prior theoretical predictions
of $T^{1/3}$ scaling. This discrepancy highlights the critical role
of gauge fluctuations in modifying low-energy transport properties.
(4) our approach correctly shows the linear-$T$ behavior of electron
specific heat at low temperatures, in agreement with earlier theoretical
predictions \citep{Werman2018,Fabrizio2023,He2023,Li2021,Zheng2024}.
(5) the effective kinetic interaction $T_{\text{bind}}$ introduced
in this approach also mimicks the resonant peaks as observed in cobalt
atoms on single-layer 1$T$-TaSe$_{2}$ \citep{Chen2022}.

The methodology presented here combines non-perturbative quantum field
theory with projective construction techniques, offering a versatile
platform to explore intertwined orders and fractionalized phases in
correlated systems. Future extensions could incorporate higher-loop
corrections to refine gauge fluctuation effects or generalize the
formalism to $Z_{2}$ or SU(2) spin liquids. Experimental verification
of our predictions---such as the $T^{-2}$-scaling thermal conductivity
and impurity-induced spectral features---would provide crucial tests
for U(1) QSL scenarios in triangular lattice materials. This work
establishes a pathway for quantitatively connecting microscopic strong
correlation physics with macroscopic observable quantities in quantum
spin liquids and related phases.

\section*{ACKNOWLEDGMENTS}

The authors would like to thank Sharif University of Technology for
supports. ZXM thanks Elahe Davari, Kosuke Nogaki, Yin Zhong and Hui
Li for helpful discussion. We also thank Microsoft's Copilot for revising
several sentences in the latest version of the manuscript to enhance
clarity.

\newpage{}

\bibliographystyle{apsrev4-2}
\bibliography{ref}

@Article{Lee2006,
  author    = {Lee, Patrick A. and Nagaosa, Naoto and Wen, Xiao-Gang},
  journal   = {Rev. Mod. Phys.},
  title     = {Doping a Mott insulator: Physics of high-temperature superconductivity},
  year      = {2006},
  month     = {Jan},
  pages     = {17--85},
  volume    = {78},
  doi       = {10.1103/RevModPhys.78.17},
  issue     = {1},
  numpages  = {0},
  publisher = {American Physical Society},
  url       = {https://link.aps.org/doi/10.1103/RevModPhys.78.17},
}

@Article{Lee2005,
  author    = {Lee, Sung-Sik and Lee, Patrick A.},
  journal   = {Phys. Rev. Lett.},
  title     = {U(1) Gauge Theory of the Hubbard Model: Spin Liquid States and Possible Application to $\ensuremath{\kappa}\mathrm{\text{\ensuremath{-}}}(\mathrm{BEDT}\mathrm{\text{\ensuremath{-}}}\mathrm{TTF}{)}_{2}{\mathrm{Cu}}_{2}(\mathrm{CN}{)}_{3}$},
  year      = {2005},
  month     = {Jul},
  pages     = {036403},
  volume    = {95},
  doi       = {10.1103/PhysRevLett.95.036403},
  issue     = {3},
  numpages  = {4},
  publisher = {American Physical Society},
  url       = {https://link.aps.org/doi/10.1103/PhysRevLett.95.036403},
}

@Article{Florens2002,
  author    = {Florens, Serge and Georges, Antoine},
  journal   = {Phys. Rev. B},
  title     = {Quantum impurity solvers using a slave rotor representation},
  year      = {2002},
  month     = {Oct},
  pages     = {165111},
  volume    = {66},
  doi       = {10.1103/PhysRevB.66.165111},
  issue     = {16},
  numpages  = {16},
  publisher = {American Physical Society},
  url       = {https://link.aps.org/doi/10.1103/PhysRevB.66.165111},
}

@Article{Florens2004,
  author    = {Florens, Serge and Georges, Antoine},
  journal   = {Phys. Rev. B},
  title     = {Slave-rotor mean-field theories of strongly correlated systems and the Mott transition in finite dimensions},
  year      = {2004},
  month     = {Jul},
  pages     = {035114},
  volume    = {70},
  doi       = {10.1103/PhysRevB.70.035114},
  issue     = {3},
  numpages  = {15},
  publisher = {American Physical Society},
  url       = {https://link.aps.org/doi/10.1103/PhysRevB.70.035114},
}

@Book{Hewson1997,
  author    = {Hewson, A.C.},
  publisher = {Cambridge University Press},
  title     = {The Kondo Problem to Heavy Fermions},
  year      = {1997},
  isbn      = {9780521599474},
  series    = {Cambridge Studies in Magnetism},
  lccn      = {98107736},
  url       = {https://books.google.com/books?id=fPzgHneNFDAC},
}

@Article{Chen2022,
  author   = {Chen, Yi and He, Wen-Yu and Ruan, Wei and Hwang, Jinwoong and Tang, Shujie and Lee, Ryan L. and Wu, Meng and Zhu, Tiancong and Zhang, Canxun and Ryu, Hyejin and Wang, Feng and Louie, Steven G. and Shen, Zhi-Xun and Mo, Sung-Kwan and Lee, Patrick A. and Crommie, Michael F.},
  journal  = {Nature Physics},
  title    = {Evidence for a spinon Kondo effect in cobalt atoms on single-layer 1T-TaSe2},
  year     = {2022},
  issn     = {1745-2481},
  number   = {11},
  pages    = {1335--1340},
  volume   = {18},
  doi      = {10.1038/s41567-022-01751-4},
  refid    = {Chen2022},
  url      = {https://doi.org/10.1038/s41567-022-01751-4},
}

@Article{He2022,
  author    = {He, Wen-Yu and Lee, Patrick A.},
  journal   = {Phys. Rev. B},
  title     = {Magnetic impurity as a local probe of the $U$(1) quantum spin liquid with spinon Fermi surface},
  year      = {2022},
  month     = {May},
  pages     = {195156},
  volume    = {105},
  doi       = {10.1103/PhysRevB.105.195156},
  issue     = {19},
  numpages  = {15},
  publisher = {American Physical Society},
  url       = {https://link.aps.org/doi/10.1103/PhysRevB.105.195156},
}

@Article{Coleman1983,
  author    = {Coleman, Piers},
  journal   = {Phys. Rev. B},
  title     = {$\frac{1}{N}$ expansion for the Kondo lattice},
  year      = {1983},
  month     = {Nov},
  pages     = {5255--5262},
  volume    = {28},
  doi       = {10.1103/PhysRevB.28.5255},
  issue     = {9},
  numpages  = {0},
  publisher = {American Physical Society},
  url       = {https://link.aps.org/doi/10.1103/PhysRevB.28.5255},
}

@Book{Coleman2015,
  author    = {Coleman, Piers},
  publisher = {Cambridge University Press},
  title     = {Introduction to Many-Body Physics},
  year      = {2015},
  place     = {Cambridge},
}

@Article{Burdin2002,
  author    = {Burdin, S. and Grempel, D. R. and Georges, A.},
  journal   = {Phys. Rev. B},
  title     = {Heavy-fermion and spin-liquid behavior in a Kondo lattice with magnetic frustration},
  year      = {2002},
  month     = {Jul},
  pages     = {045111},
  volume    = {66},
  doi       = {10.1103/PhysRevB.66.045111},
  issue     = {4},
  numpages  = {6},
  publisher = {American Physical Society},
  url       = {https://link.aps.org/doi/10.1103/PhysRevB.66.045111},
}

@Book{Sachdev2023,
  author    = {Sachdev, Subir},
  publisher = {Cambridge University Press},
  title     = {Quantum Phases of Matter},
  year      = {2023},
  place     = {Cambridge},
}

@Article{Law2017,
  author   = {K. T. Law and Patrick A. Lee},
  journal  = {Proceedings of the National Academy of Sciences},
  title    = {1T-TaS<sub>2</sub> as a quantum spin liquid},
  year     = {2017},
  number   = {27},
  pages    = {6996-7000},
  volume   = {114},
  doi      = {10.1073/pnas.1706769114},
  eprint   = {https://www.pnas.org/doi/pdf/10.1073/pnas.1706769114},
  url      = {https://www.pnas.org/doi/abs/10.1073/pnas.1706769114},
}

@Article{Lee2009,
  author    = {Lee, Sung-Sik},
  journal   = {Phys. Rev. B},
  title     = {Low-energy effective theory of Fermi surface coupled with U(1) gauge field in $2+1$ dimensions},
  year      = {2009},
  month     = {Oct},
  pages     = {165102},
  volume    = {80},
  doi       = {10.1103/PhysRevB.80.165102},
  issue     = {16},
  numpages  = {13},
  publisher = {American Physical Society},
  url       = {https://link.aps.org/doi/10.1103/PhysRevB.80.165102},
}

@Article{Podolsky2009,
  author    = {Podolsky, Daniel and Paramekanti, Arun and Kim, Yong Baek and Senthil, T.},
  journal   = {Phys. Rev. Lett.},
  title     = {Mott Transition between a Spin-Liquid Insulator and a Metal in Three Dimensions},
  year      = {2009},
  month     = {May},
  pages     = {186401},
  volume    = {102},
  doi       = {10.1103/PhysRevLett.102.186401},
  issue     = {18},
  numpages  = {4},
  publisher = {American Physical Society},
  url       = {https://link.aps.org/doi/10.1103/PhysRevLett.102.186401},
}

@Article{Hermele2004,
  author    = {Hermele, Michael and Senthil, T. and Fisher, Matthew P. A. and Lee, Patrick A. and Nagaosa, Naoto and Wen, Xiao-Gang},
  journal   = {Phys. Rev. B},
  title     = {Stability of $U(1)$ spin liquids in two dimensions},
  year      = {2004},
  month     = {Dec},
  pages     = {214437},
  volume    = {70},
  doi       = {10.1103/PhysRevB.70.214437},
  issue     = {21},
  numpages  = {9},
  publisher = {American Physical Society},
  url       = {https://link.aps.org/doi/10.1103/PhysRevB.70.214437},
}

@Article{Senthil2008,
  author    = {Senthil, T.},
  journal   = {Phys. Rev. B},
  title     = {Theory of a continuous Mott transition in two dimensions},
  year      = {2008},
  month     = {Jul},
  pages     = {045109},
  volume    = {78},
  doi       = {10.1103/PhysRevB.78.045109},
  issue     = {4},
  numpages  = {13},
  publisher = {American Physical Society},
  url       = {https://link.aps.org/doi/10.1103/PhysRevB.78.045109},
}

@Article{Lee2008,
  author    = {Lee, Sung-Sik},
  journal   = {Phys. Rev. B},
  title     = {Stability of the U(1) spin liquid with a spinon Fermi surface in $2+1$ dimensions},
  year      = {2008},
  month     = {Aug},
  pages     = {085129},
  volume    = {78},
  doi       = {10.1103/PhysRevB.78.085129},
  issue     = {8},
  numpages  = {13},
  publisher = {American Physical Society},
  url       = {https://link.aps.org/doi/10.1103/PhysRevB.78.085129},
}

@Article{Nogaki2023,
  author    = {Kosuke Nogaki and Jiani Fei and Emanuel Gull and Hiroshi Shinaoka},
  journal   = {SciPost Phys. Codebases},
  title     = {{Nevanlinna.jl: A Julia implementation of Nevanlinna analytic continuation}},
  year      = {2023},
  pages     = {19},
  doi       = {10.21468/SciPostPhysCodeb.19},
  publisher = {SciPost},
  url       = {https://scipost.org/10.21468/SciPostPhysCodeb.19},
}

@Article{Fei2021,
  author    = {Fei, Jiani and Yeh, Chia-Nan and Gull, Emanuel},
  journal   = {Phys. Rev. Lett.},
  title     = {Nevanlinna Analytical Continuation},
  year      = {2021},
  month     = {Feb},
  pages     = {056402},
  volume    = {126},
  doi       = {10.1103/PhysRevLett.126.056402},
  issue     = {5},
  numpages  = {6},
  publisher = {American Physical Society},
  url       = {https://link.aps.org/doi/10.1103/PhysRevLett.126.056402},
}

@Article{Nogaki2023a,
  author   = {Nogaki ,Kosuke and Shinaoka ,Hiroshi},
  journal  = {Journal of the Physical Society of Japan},
  title    = {Bosonic Nevanlinna Analytic Continuation},
  year     = {2023},
  number   = {3},
  pages    = {035001},
  volume   = {92},
  doi      = {10.7566/JPSJ.92.035001},
  eprint   = {https://doi.org/10.7566/JPSJ.92.035001},
  url      = {https://doi.org/10.7566/JPSJ.92.035001},
}

@Article{Shinaoka2022,
  author    = {Hiroshi Shinaoka and Naoya Chikano and Emanuel Gull and Jia Li and Takuya Nomoto and Junya Otsuki and Markus Wallerberger and Tianchun Wang and Kazuyoshi Yoshimi},
  journal   = {SciPost Phys. Lect. Notes},
  title     = {{Efficient ab initio many-body calculations based on sparse modeling of Matsubara Green's function}},
  year      = {2022},
  pages     = {63},
  doi       = {10.21468/SciPostPhysLectNotes.63},
  publisher = {SciPost},
  url       = {https://scipost.org/10.21468/SciPostPhysLectNotes.63},
}

@Article{Shinaoka2017,
  author    = {Shinaoka, Hiroshi and Otsuki, Junya and Ohzeki, Masayuki and Yoshimi, Kazuyoshi},
  journal   = {Phys. Rev. B},
  title     = {Compressing Green's function using intermediate representation between imaginary-time and real-frequency domains},
  year      = {2017},
  month     = {Jul},
  pages     = {035147},
  volume    = {96},
  doi       = {10.1103/PhysRevB.96.035147},
  issue     = {3},
  numpages  = {8},
  publisher = {American Physical Society},
  url       = {https://link.aps.org/doi/10.1103/PhysRevB.96.035147},
}

@Article{Otsuki2020,
  author   = {Otsuki ,Junya and Ohzeki ,Masayuki and Shinaoka ,Hiroshi and Yoshimi ,Kazuyoshi},
  journal  = {Journal of the Physical Society of Japan},
  title    = {Sparse Modeling in Quantum Many-Body Problems},
  year     = {2020},
  number   = {1},
  pages    = {012001},
  volume   = {89},
  doi      = {10.7566/JPSJ.89.012001},
  eprint   = {https://doi.org/10.7566/JPSJ.89.012001},
  url      = {https://doi.org/10.7566/JPSJ.89.012001},
}

@Article{Luttinger1960,
  author    = {Luttinger, J. M. and Ward, J. C.},
  journal   = {Phys. Rev.},
  title     = {Ground-State Energy of a Many-Fermion System. II},
  year      = {1960},
  month     = {Jun},
  pages     = {1417--1427},
  volume    = {118},
  doi       = {10.1103/PhysRev.118.1417},
  issue     = {5},
  numpages  = {0},
  publisher = {American Physical Society},
  url       = {https://link.aps.org/doi/10.1103/PhysRev.118.1417},
}

@Article{Baym1961,
  author    = {Baym, Gordon and Kadanoff, Leo P.},
  journal   = {Phys. Rev.},
  title     = {Conservation Laws and Correlation Functions},
  year      = {1961},
  month     = {Oct},
  pages     = {287--299},
  volume    = {124},
  doi       = {10.1103/PhysRev.124.287},
  issue     = {2},
  numpages  = {0},
  publisher = {American Physical Society},
  url       = {https://link.aps.org/doi/10.1103/PhysRev.124.287},
}

@Article{Baym1962,
  author    = {Baym, Gordon},
  journal   = {Phys. Rev.},
  title     = {Self-Consistent Approximations in Many-Body Systems},
  year      = {1962},
  month     = {Aug},
  pages     = {1391--1401},
  volume    = {127},
  doi       = {10.1103/PhysRev.127.1391},
  issue     = {4},
  numpages  = {0},
  publisher = {American Physical Society},
  url       = {https://link.aps.org/doi/10.1103/PhysRev.127.1391},
}

@Book{Stefanucci2025,
  author    = {Stefanucci, Gianluca and van Leeuwen, Robert},
  publisher = {Cambridge University Press},
  title     = {Nonequilibrium Many-Body Theory of Quantum Systems: A Modern Introduction},
  year      = {2025},
  edition   = {2},
  place     = {Cambridge},
}

@Article{Sumita2023,
  author    = {Sumita, Shuntaro and Naka, Makoto and Seo, Hitoshi},
  journal   = {Phys. Rev. Res.},
  title     = {Fulde-Ferrell-Larkin-Ovchinnikov state induced by antiferromagnetic order in $\ensuremath{\kappa}$-type organic conductors},
  year      = {2023},
  month     = {Nov},
  pages     = {043171},
  volume    = {5},
  doi       = {10.1103/PhysRevResearch.5.043171},
  issue     = {4},
  numpages  = {17},
  publisher = {American Physical Society},
  url       = {https://link.aps.org/doi/10.1103/PhysRevResearch.5.043171},
}

@Article{Witt2021,
  author    = {Witt, Niklas and van Loon, Erik G. C. P. and Nomoto, Takuya and Arita, Ryotaro and Wehling, Tim O.},
  journal   = {Phys. Rev. B},
  title     = {Efficient fluctuation-exchange approach to low-temperature spin fluctuations and superconductivity: From the Hubbard model to ${\mathrm{Na}}_{x}{\mathrm{CoO}}_{2}\ifmmode\cdot\else\textperiodcentered\fi{}{y\mathrm{H}}_{2}\mathrm{O}$},
  year      = {2021},
  month     = {May},
  pages     = {205148},
  volume    = {103},
  doi       = {10.1103/PhysRevB.103.205148},
  issue     = {20},
  numpages  = {12},
  publisher = {American Physical Society},
  url       = {https://link.aps.org/doi/10.1103/PhysRevB.103.205148},
}

@PhdThesis{Li2023,
  author = {Li, Hui},
  school = {Peking University},
  title  = {Quantum Many-Body Self-Consistent Theory and Its Applications},
  year   = {2023},
}

@Article{Quinn1958,
  author    = {Quinn, John J. and Ferrell, Richard A.},
  journal   = {Phys. Rev.},
  title     = {Electron Self-Energy Approach to Correlation in a Degenerate Electron Gas},
  year      = {1958},
  month     = {Nov},
  pages     = {812--827},
  volume    = {112},
  doi       = {10.1103/PhysRev.112.812},
  issue     = {3},
  numpages  = {0},
  publisher = {American Physical Society},
  url       = {https://link.aps.org/doi/10.1103/PhysRev.112.812},
}

@Article{Hedin1965,
  author    = {Hedin, Lars},
  journal   = {Phys. Rev.},
  title     = {New Method for Calculating the One-Particle Green's Function with Application to the Electron-Gas Problem},
  year      = {1965},
  month     = {Aug},
  pages     = {A796--A823},
  volume    = {139},
  doi       = {10.1103/PhysRev.139.A796},
  issue     = {3A},
  numpages  = {0},
  publisher = {American Physical Society},
  url       = {https://link.aps.org/doi/10.1103/PhysRev.139.A796},
}

@Article{Aryasetiawan1998,
  author   = {F Aryasetiawan and O Gunnarsson},
  journal  = {Reports on Progress in Physics},
  title    = {The GW method},
  year     = {1998},
  month    = {mar},
  number   = {3},
  pages    = {237},
  volume   = {61},
  doi      = {10.1088/0034-4885/61/3/002},
  url      = {https://dx.doi.org/10.1088/0034-4885/61/3/002},
}

@Misc{Li2024,
  author        = {Hui Li and Yingze Su and Junnian Xiong and Haiqing Lin and Huaqing Huang and Dingping Li},
  title         = {Post-$GW$ theory and its application to pseudogap in strongly correlated system},
  year          = {2024},
  archiveprefix = {arXiv},
  eprint        = {2409.16762},
  primaryclass  = {cond-mat.str-el},
  url           = {https://arxiv.org/abs/2409.16762},
}

@Article{Sun2021,
  author    = {Sun, Zhipeng and Fan, Zhenhao and Li, Hui and Li, Dingping and Rosenstein, Baruch},
  journal   = {Phys. Rev. B},
  title     = {Modified $GW$ method in electronic systems},
  year      = {2021},
  month     = {Sep},
  pages     = {125137},
  volume    = {104},
  doi       = {10.1103/PhysRevB.104.125137},
  issue     = {12},
  numpages  = {15},
  publisher = {American Physical Society},
  url       = {https://link.aps.org/doi/10.1103/PhysRevB.104.125137},
}

@Article{Zheng2024,
  author    = {Zheng, Xia-Ming and Kargarian, Mehdi},
  journal   = {Phys. Rev. B},
  title     = {Spinon Kondo lattice in quantum spin liquids using the slave-rotor formalism},
  year      = {2024},
  month     = {Sep},
  pages     = {115116},
  volume    = {110},
  doi       = {10.1103/PhysRevB.110.115116},
  issue     = {11},
  numpages  = {11},
  publisher = {American Physical Society},
  url       = {https://link.aps.org/doi/10.1103/PhysRevB.110.115116},
}

@Book{Schwinger2015,
  author    = {Schwinger, J.},
  publisher = {Dover Publications},
  title     = {On Angular Momentum},
  year      = {2015},
  isbn      = {9780486788104},
  series    = {Dover Books on Physics},
  lccn      = {2014036399},
  url       = {https://books.google.com/books?id=78m3BQAAQBAJ},
}

@Article{Abrikosov1965,
  author    = {Abrikosov, A. A.},
  journal   = {Physics Physique Fizika},
  title     = {Electron scattering on magnetic impurities in metals and anomalous resistivity effects},
  year      = {1965},
  month     = {Sep},
  pages     = {5--20},
  volume    = {2},
  doi       = {10.1103/PhysicsPhysiqueFizika.2.5},
  issue     = {1},
  numpages  = {16},
  publisher = {American Physical Society},
  url       = {https://link.aps.org/doi/10.1103/PhysicsPhysiqueFizika.2.5},
}

@Article{Christos2023,
  author    = {Christos, Maine and Luo, Zhu-Xi and Shackleton, Henry and Zhang, Ya-Hui and Scheurer, Mathias S. and Sachdev, Subir},
  journal   = {Proceedings of the National Academy of Sciences},
  title     = {A model ofd-wave superconductivity, antiferromagnetism, and charge order on the square lattice},
  year      = {2023},
  issn      = {1091-6490},
  month     = may,
  number    = {21},
  volume    = {120},
  doi       = {10.1073/pnas.2302701120},
  publisher = {Proceedings of the National Academy of Sciences},
}

@Book{Dupuis2023,
  author    = {Dupuis, N.},
  publisher = {World Scientific Publishing Europe Limited},
  title     = {Field Theory Of Condensed Matter And Ultracold Gases - Volume 1},
  year      = {2023},
  isbn      = {9781800613911},
  url       = {https://books.google.com/books?id=b0Ex0AEACAAJ},
}

@Article{Balents2020,
  author    = {Balents, Leon and Starykh, Oleg A.},
  journal   = {Phys. Rev. B},
  title     = {Collective spinon spin wave in a magnetized U(1) spin liquid},
  year      = {2020},
  month     = {Jan},
  pages     = {020401},
  volume    = {101},
  doi       = {10.1103/PhysRevB.101.020401},
  issue     = {2},
  numpages  = {5},
  publisher = {American Physical Society},
  url       = {https://link.aps.org/doi/10.1103/PhysRevB.101.020401},
}

@Article{Li2020,
  author    = {Li, Jia and Wallerberger, Markus and Chikano, Naoya and Yeh, Chia-Nan and Gull, Emanuel and Shinaoka, Hiroshi},
  journal   = {Phys. Rev. B},
  title     = {Sparse sampling approach to efficient ab initio calculations at finite temperature},
  year      = {2020},
  month     = {Jan},
  pages     = {035144},
  volume    = {101},
  doi       = {10.1103/PhysRevB.101.035144},
  issue     = {3},
  numpages  = {12},
  publisher = {American Physical Society},
  url       = {https://link.aps.org/doi/10.1103/PhysRevB.101.035144},
}

@Article{Werman2018,
  author    = {Werman, Yochai and Chatterjee, Shubhayu and Morampudi, Siddhardh C. and Berg, Erez},
  journal   = {Phys. Rev. X},
  title     = {Signatures of Fractionalization in Spin Liquids from Interlayer Thermal Transport},
  year      = {2018},
  month     = {Sep},
  pages     = {031064},
  volume    = {8},
  doi       = {10.1103/PhysRevX.8.031064},
  issue     = {3},
  numpages  = {23},
  publisher = {American Physical Society},
  url       = {https://link.aps.org/doi/10.1103/PhysRevX.8.031064},
}

@Article{Scheurer2018,
  author    = {Scheurer, Mathias S. and Sachdev, Subir},
  journal   = {Phys. Rev. B},
  title     = {Orbital currents in insulating and doped antiferromagnets},
  year      = {2018},
  month     = {Dec},
  pages     = {235126},
  volume    = {98},
  doi       = {10.1103/PhysRevB.98.235126},
  issue     = {23},
  numpages  = {22},
  publisher = {American Physical Society},
  url       = {https://link.aps.org/doi/10.1103/PhysRevB.98.235126},
}

@Article{Khatibi2020,
  author    = {Khatibi, Zahra and Ahemeh, Roya and Kargarian, Mehdi},
  journal   = {Phys. Rev. B},
  title     = {Excitonic insulator phase and condensate dynamics in a topological one-dimensional model},
  year      = {2020},
  month     = {Dec},
  pages     = {245121},
  volume    = {102},
  doi       = {10.1103/PhysRevB.102.245121},
  issue     = {24},
  numpages  = {12},
  publisher = {American Physical Society},
  url       = {https://link.aps.org/doi/10.1103/PhysRevB.102.245121},
}

@Article{Fabrizio2023,
  author    = {Fabrizio, Michele},
  journal   = {Phys. Rev. Lett.},
  title     = {Spin-Liquid Insulators Can Be Landau's Fermi Liquids},
  year      = {2023},
  month     = {Apr},
  pages     = {156702},
  volume    = {130},
  doi       = {10.1103/PhysRevLett.130.156702},
  issue     = {15},
  numpages  = {6},
  publisher = {American Physical Society},
  url       = {https://link.aps.org/doi/10.1103/PhysRevLett.130.156702},
}

@Article{Dee2019,
  author    = {Dee, P. M. and Nakatsukasa, K. and Wang, Y. and Johnston, S.},
  journal   = {Phys. Rev. B},
  title     = {Temperature-filling phase diagram of the two-dimensional Holstein model in the thermodynamic limit by self-consistent Migdal approximation},
  year      = {2019},
  month     = {Jan},
  pages     = {024514},
  volume    = {99},
  doi       = {10.1103/PhysRevB.99.024514},
  issue     = {2},
  numpages  = {18},
  publisher = {American Physical Society},
  url       = {https://link.aps.org/doi/10.1103/PhysRevB.99.024514},
}

@Article{Zhang2020,
  author    = {Zhang, Ya-Hui and Sachdev, Subir},
  journal   = {Phys. Rev. Res.},
  title     = {From the pseudogap metal to the Fermi liquid using ancilla qubits},
  year      = {2020},
  month     = {May},
  pages     = {023172},
  volume    = {2},
  doi       = {10.1103/PhysRevResearch.2.023172},
  issue     = {2},
  numpages  = {8},
  publisher = {American Physical Society},
  url       = {https://link.aps.org/doi/10.1103/PhysRevResearch.2.023172},
}

@Article{Nave2007,
  author    = {Nave, Cody P. and Lee, Patrick A.},
  journal   = {Phys. Rev. B},
  title     = {Transport properties of a spinon Fermi surface coupled to a U(1) gauge field},
  year      = {2007},
  month     = {Dec},
  pages     = {235124},
  volume    = {76},
  doi       = {10.1103/PhysRevB.76.235124},
  issue     = {23},
  numpages  = {11},
  publisher = {American Physical Society},
  url       = {https://link.aps.org/doi/10.1103/PhysRevB.76.235124},
}

@Misc{He2023,
  author        = {He, Wenyu},
  title         = {Quantum Spin Liquid Physics in 1T-TaS2 and 1T-TaSe2},
  year          = {2023},
  archiveprefix = {koushare},
  url           = {https://www.koushare.com/live/details/18687},
}

@Article{Li2021,
  author    = {Li, Tao},
  journal   = {Phys. Rev. B},
  title     = {Absence of a ${T}^{2/3}$ specific heat anomaly in a $U(1)$ spin liquid with a large spinon Fermi surface},
  year      = {2021},
  month     = {Oct},
  pages     = {165123},
  volume    = {104},
  doi       = {10.1103/PhysRevB.104.165123},
  issue     = {16},
  numpages  = {11},
  publisher = {American Physical Society},
  url       = {https://link.aps.org/doi/10.1103/PhysRevB.104.165123},
}

@Article{Lee1992,
  author    = {Lee, Patrick A. and Nagaosa, Naoto},
  journal   = {Phys. Rev. B},
  title     = {Gauge theory of the normal state of high-${\mathit{T}}_{\mathit{c}}$ superconductors},
  year      = {1992},
  month     = {Sep},
  pages     = {5621--5639},
  volume    = {46},
  doi       = {10.1103/PhysRevB.46.5621},
  issue     = {9},
  numpages  = {0},
  publisher = {American Physical Society},
  url       = {https://link.aps.org/doi/10.1103/PhysRevB.46.5621},
}

@Article{Nagaosa1990,
  author    = {Nagaosa, Naoto and Lee, Patrick A.},
  journal   = {Phys. Rev. Lett.},
  title     = {Normal-state properties of the uniform resonating-valence-bond state},
  year      = {1990},
  month     = {May},
  pages     = {2450--2453},
  volume    = {64},
  doi       = {10.1103/PhysRevLett.64.2450},
  issue     = {20},
  numpages  = {0},
  publisher = {American Physical Society},
  url       = {https://link.aps.org/doi/10.1103/PhysRevLett.64.2450},
}

@Article{He2023a,
  author    = {He, Wen-Yu and Lee, Patrick A.},
  journal   = {Phys. Rev. B},
  title     = {Electronic density of states of a $U(1)$ quantum spin liquid with spinon Fermi surface. I. Orbital magnetic field effects},
  year      = {2023},
  month     = {May},
  pages     = {195155},
  volume    = {107},
  doi       = {10.1103/PhysRevB.107.195155},
  issue     = {19},
  numpages  = {16},
  publisher = {American Physical Society},
  url       = {https://link.aps.org/doi/10.1103/PhysRevB.107.195155},
}

@Book{Montvay1994,
  author    = {Montvay, I. and M{\"u}nster, G.},
  publisher = {Cambridge University Press},
  title     = {Quantum Fields on a Lattice},
  year      = {1994},
  isbn      = {9780521599177},
  series    = {Cambridge Monographs on Mathematical Physics},
  lccn      = {93001026},
  url       = {https://books.google.com.hk/books?id=NHZshmEBXhcC},
}

@Article{Gunnarsson2010,
  author    = {Gunnarsson, O. and Haverkort, M. W. and Sangiovanni, G.},
  journal   = {Phys. Rev. B},
  title     = {Analytical continuation of imaginary axis data for optical conductivity},
  year      = {2010},
  month     = {Oct},
  pages     = {165125},
  volume    = {82},
  doi       = {10.1103/PhysRevB.82.165125},
  issue     = {16},
  numpages  = {8},
  publisher = {American Physical Society},
  url       = {https://link.aps.org/doi/10.1103/PhysRevB.82.165125},
}

@Article{Huang2023,
  author   = {Li Huang},
  journal  = {Computer Physics Communications},
  title    = {ACFlow: An open source toolkit for analytic continuation of quantum Monte Carlo data},
  year     = {2023},
  issn     = {0010-4655},
  pages    = {108863},
  volume   = {292},
  doi      = {https://doi.org/10.1016/j.cpc.2023.108863},
  url      = {https://www.sciencedirect.com/science/article/pii/S0010465523002084},
}

@Article{Mei2022,
  author   = {Mei, Hongying and Yuan, Haifeng and Wen, Hua and Yao, Haizi and Sun, Shuxiang and Zheng, Xinyan and Liu, Fang and Li, Haowen and Xu, Wen},
  journal  = {The European Physical Journal B},
  title    = {Optical conductivity of an electron gas driven by a pulsed terahertz radiation field},
  year     = {2022},
  issn     = {1434-6036},
  number   = {7},
  pages    = {111},
  volume   = {95},
  doi      = {10.1140/epjb/s10051-022-00363-4},
  refid    = {Mei2022},
  url      = {https://doi.org/10.1140/epjb/s10051-022-00363-4},
}

@Article{Blaizot2021,
  author   = {Jean-Paul Blaizot and Jan M. Pawlowski and Urko Reinosa},
  journal  = {Annals of Physics},
  title    = {Functional renormalization group and 2PI effective action formalism},
  year     = {2021},
  issn     = {0003-4916},
  pages    = {168549},
  volume   = {431},
  doi      = {https://doi.org/10.1016/j.aop.2021.168549},
  url      = {https://www.sciencedirect.com/science/article/pii/S000349162100155X},
}

@Article{Yanase2005,
  author   = {Yanase ,Youichi and Ogata ,Masao},
  journal  = {Journal of the Physical Society of Japan},
  title    = {Kinetic Energy, Condensation Energy, Optical Sum Rule and Pairing Mechanism in High-Tc Cuprates},
  year     = {2005},
  number   = {5},
  pages    = {1534-1543},
  volume   = {74},
  doi      = {10.1143/JPSJ.74.1534},
  eprint   = {https://doi.org/10.1143/JPSJ.74.1534},
  url      = {https://doi.org/10.1143/JPSJ.74.1534},
}

@Book{Fetter2003,
  author    = {Fetter, A.L. and Walecka, J.D.},
  publisher = {Dover Publications},
  title     = {Quantum Theory of Many-particle Systems},
  year      = {2003},
  isbn      = {9780486428277},
  series    = {Dover Books on Physics},
  lccn      = {2003043536},
  url       = {https://books.google.com/books?id=0wekf1s83b0C},
}

@InBook{Mahan2000,
  author    = {Mahan, Gerald D.},
  publisher = {Springer US},
  year      = {2000},
  address   = {Boston, MA},
  isbn      = {978-1-4757-5714-9},
  booktitle = {Many-Particle Physics},
  doi       = {10.1007/978-1-4757-5714-9},
  url       = {https://doi.org/10.1007/978-1-4757-5714-9},
}

@Article{Huang2019,
  author   = {Edwin W. Huang and Ryan Sheppard and Brian Moritz and Thomas P. Devereaux},
  journal  = {Science},
  title    = {Strange metallicity in the doped Hubbard model},
  year     = {2019},
  number   = {6468},
  pages    = {987-990},
  volume   = {366},
  doi      = {10.1126/science.aau7063},
  eprint   = {https://www.science.org/doi/pdf/10.1126/science.aau7063},
  url      = {https://www.science.org/doi/abs/10.1126/science.aau7063},
}

@Article{Shirakawa2017,
  author    = {Shirakawa, Tomonori and Tohyama, Takami and Kokalj, Jure and Sota, Sigetoshi and Yunoki, Seiji},
  journal   = {Phys. Rev. B},
  title     = {Ground-state phase diagram of the triangular lattice Hubbard model by the density-matrix renormalization group method},
  year      = {2017},
  month     = {Nov},
  pages     = {205130},
  volume    = {96},
  doi       = {10.1103/PhysRevB.96.205130},
  issue     = {20},
  numpages  = {11},
  publisher = {American Physical Society},
  url       = {https://link.aps.org/doi/10.1103/PhysRevB.96.205130},
}

@Article{Yamada2014,
  author    = {Yamada, A.},
  journal   = {Phys. Rev. B},
  title     = {Magnetic properties and Mott transition in the Hubbard model on the anisotropic triangular lattice},
  year      = {2014},
  month     = {May},
  pages     = {195108},
  volume    = {89},
  doi       = {10.1103/PhysRevB.89.195108},
  issue     = {19},
  numpages  = {9},
  publisher = {American Physical Society},
  url       = {https://link.aps.org/doi/10.1103/PhysRevB.89.195108},
}

@Article{Sahebsara2008,
  author    = {Sahebsara, Peyman and S\'en\'echal, David},
  journal   = {Phys. Rev. Lett.},
  title     = {Hubbard Model on the Triangular Lattice: Spiral Order and Spin Liquid},
  year      = {2008},
  month     = {Mar},
  pages     = {136402},
  volume    = {100},
  doi       = {10.1103/PhysRevLett.100.136402},
  issue     = {13},
  numpages  = {4},
  publisher = {American Physical Society},
  url       = {https://link.aps.org/doi/10.1103/PhysRevLett.100.136402},
}

@Article{Clay2008,
  author    = {Clay, R. T. and Li, H. and Mazumdar, S.},
  journal   = {Phys. Rev. Lett.},
  title     = {Absence of Superconductivity in the Half-Filled Band Hubbard Model on the Anisotropic Triangular Lattice},
  year      = {2008},
  month     = {Oct},
  pages     = {166403},
  volume    = {101},
  doi       = {10.1103/PhysRevLett.101.166403},
  issue     = {16},
  numpages  = {4},
  publisher = {American Physical Society},
  url       = {https://link.aps.org/doi/10.1103/PhysRevLett.101.166403},
}

@Article{Kokalj2013,
  author    = {Kokalj, J. and McKenzie, Ross H.},
  journal   = {Phys. Rev. Lett.},
  title     = {Thermodynamics of a Bad Metal--Mott Insulator Transition in the Presence of Frustration},
  year      = {2013},
  month     = {May},
  pages     = {206402},
  volume    = {110},
  doi       = {10.1103/PhysRevLett.110.206402},
  issue     = {20},
  numpages  = {5},
  publisher = {American Physical Society},
  url       = {https://link.aps.org/doi/10.1103/PhysRevLett.110.206402},
}

@Article{Szasz2020,
  author    = {Szasz, Aaron and Motruk, Johannes and Zaletel, Michael P. and Moore, Joel E.},
  journal   = {Phys. Rev. X},
  title     = {Chiral Spin Liquid Phase of the Triangular Lattice Hubbard Model: A Density Matrix Renormalization Group Study},
  year      = {2020},
  month     = {May},
  pages     = {021042},
  volume    = {10},
  doi       = {10.1103/PhysRevX.10.021042},
  issue     = {2},
  numpages  = {16},
  publisher = {American Physical Society},
  url       = {https://link.aps.org/doi/10.1103/PhysRevX.10.021042},
}

\cleardoublepage{}

\appendix

\section{Current, Energy Current, and Heat Current Density Operators and Their
Correlation Functions on the Lattice\label{Appendix Current operator}}
\begin{widetext}
To calculate the thermal conductivity of spinons, it is essential
to consider the correlation function of the heat current density.
In this appendix, we shall present, in sequential order, the expressions
for various current densities on the lattice, culminating in an exposition
of the Green's function methodology for computing the heat-current
correlation function.

The particle current density, which is proportional to momentum operator,
is expressed as \citep{Montvay1994,Mahan2000}:

\begin{equation}
\boldsymbol{j}_{i}(\boldsymbol{r})=\frac{1}{ima}\sum_{i}\cos(\phi_{i})(\psi^{\dagger}(\boldsymbol{r})\psi(\boldsymbol{r}+\boldsymbol{a}_{i})-\psi^{\dagger}(\boldsymbol{r}+\boldsymbol{a}_{i})\psi(\boldsymbol{r})),
\end{equation}
where $\boldsymbol{a}_{i}$ represents the lattice vector. The term
$\cos(\phi_{i})$ represents the cosine of the angle between the lattice
vector and the direction of the energy flow (in this case, evidently
the $x$-direction). The angle $\phi_{i}$ is defined as $\phi_{i}=\arctan(a_{1,y}/a_{1,x})$.
By Fourier transforming the energy flow density into the crystal momentum
representation, the $x$-component becomes:

\begin{equation}
j_{x}(\boldsymbol{q})=\frac{1}{N}\sum_{\boldsymbol{k}}\left[\frac{1}{2}\left(\left.\frac{\partial E(\boldsymbol{k})}{\partial k_{x}}\right|_{\boldsymbol{k}}+\left.\frac{\partial E(\boldsymbol{k})}{\partial k_{x}}\right|_{\boldsymbol{k}+\boldsymbol{q}}\right)\psi^{\dagger}(\boldsymbol{k})\psi(\boldsymbol{k}+\boldsymbol{q})\right],
\end{equation}
where the operator $\psi$ is any Schrodinger bosonic or fermionic
operator and $E(\boldsymbol{k})$ denotes the particle energy. In
the non-interacting condition, the energy naturally reduces to the
lattice structure factor $E(\boldsymbol{k})=-t\gamma(\boldsymbol{k})$.

Here, only the $xx$-component of the correlation function is calculated:
$\Pi_{xx}^{(p)}(x_{2},x_{1})=-\left\langle j_{x}(\boldsymbol{r}_{2},\tau_{2})j_{x}(\boldsymbol{r}_{1},\tau_{1})\right\rangle $.
where $x=(\boldsymbol{r},\tau)$ is the Euclidean spacetime coordinate.
Substituting the $j_{x}$ operator into $\Pi_{xx}^{(p)}$ , the expansion
consists of multiple ensemble averages of products of four $\psi$-operators,
which are:

\begin{align}
\left\langle \psi^{\dagger}(x_{2})\psi(x_{2}+\boldsymbol{a}_{i})\psi^{\dagger}(x_{1})\psi(x_{1}+\boldsymbol{a}_{j})\right\rangle = & \pm G_{\psi}(x_{2}+\boldsymbol{a}_{i},x_{1})G_{\psi}(r_{1}+\boldsymbol{a}_{j},x_{2})=\pm G_{\psi}(\Delta x+\boldsymbol{a}_{i})G_{\psi}(-\Delta x_{1}+\boldsymbol{a}_{j})\nonumber \\
\left\langle \psi^{\dagger}(x_{2}+\boldsymbol{a}_{i})\psi(x_{2})\psi^{\dagger}(x_{1})\psi(x_{1}+\boldsymbol{a}_{j})\right\rangle = & \pm G_{\psi}(x_{2},x_{1})G_{\psi}(x_{1}+\boldsymbol{a}_{j},x_{2}+\boldsymbol{a}_{i})=\pm G_{\psi}(\Delta x)G_{\psi}(-\Delta x+\Delta\boldsymbol{a}_{ji})\nonumber \\
\left\langle \psi^{\dagger}(x_{2})\psi(x_{2}+\boldsymbol{a}_{i})\psi^{\dagger}(x_{1}+\boldsymbol{a}_{j})\psi(x_{1})\right\rangle = & \pm G_{\psi}(x_{2}+\boldsymbol{a}_{i},x_{1}+\boldsymbol{a}_{j})G_{\psi}(x_{1},x_{2})=\pm G_{\psi}(\Delta x+\Delta\boldsymbol{a}_{ij})G_{\psi}(-\Delta x)\nonumber \\
\left\langle \psi^{\dagger}(x_{2}+\boldsymbol{a}_{i})\psi(x_{2})\psi^{\dagger}(x_{1}+\boldsymbol{a}_{j})\psi(x_{1})\right\rangle = & \pm G_{\psi}(x_{2},x_{1}+\boldsymbol{a}_{j})G_{\psi}(x_{1},x_{2}+\boldsymbol{a}_{i})=\pm G_{\psi}(\Delta x-\boldsymbol{a}_{j})G_{\psi}(-\Delta x-\boldsymbol{a}_{i})
\end{align}
where $\pm$ takes $-1$ for fermions and $+1$ for bosons, and $G_{\psi}(x_{2},x_{1})$
denotes the single-particle Green's function corresponding to the
operator $\psi$ . In the case of equilibrium systems with lattice
symmetry, the Green's function depends only on $\Delta x=x_{2}-x_{2}=(\boldsymbol{r}_{2}-\boldsymbol{r}_{1},\tau_{2}-\tau_{1})$.
The symbol $\Delta\boldsymbol{a}_{ij}=\boldsymbol{a}_{i}-\boldsymbol{a}_{j}$
is definitely as the difference between two lattice vector. Consequently,
the spatial-temporal correlation function becomes:

\begin{align}
\Pi_{xx}^{(p)}(\Delta x)= & \pm\frac{1}{a^{2}}\sum_{ij}\cos(\phi_{i})\cos(\phi_{j})\left[G_{\psi}(\Delta x+\boldsymbol{a}_{i})G_{\psi}(-\Delta x_{1}+\boldsymbol{a}_{j})+G_{\psi}(\Delta x)G_{\psi}(-\Delta x+\Delta\boldsymbol{a}_{ji})\right.\nonumber \\
 & \left.+G_{\psi}(\Delta x+\Delta\boldsymbol{a}_{ij})G_{\psi}(-\Delta x)+G_{\psi}(\Delta x-\boldsymbol{a}_{j})G_{\psi}(-\Delta x-\boldsymbol{a}_{i})\right]
\end{align}

After Fourier transformation, the momentum-Matsubara frequency correlation
function is obtained:

\begin{equation}
\Pi_{xx}^{(p)}(k)=\pm\sum_{k}\Gamma_{x}(\boldsymbol{k},\boldsymbol{k}+\boldsymbol{q})\Gamma_{x}(\boldsymbol{k}+\boldsymbol{q},\boldsymbol{k})G_{\psi}(\boldsymbol{k},i\nu_{n})G_{\psi}(\boldsymbol{k}+\boldsymbol{q},i\nu_{n}+i\omega_{n}),\label{eq:A particle current correlation function}
\end{equation}
where the bare current vertex is defined as: $\Gamma_{x}^{(p)}(\boldsymbol{k},\boldsymbol{k}+\boldsymbol{q})=\frac{1}{2}\left(\left.\frac{\partial E}{\partial k_{x}}\right|_{\boldsymbol{k}}+\left.\frac{\partial E}{\partial k_{x}}\right|_{\boldsymbol{k}+\boldsymbol{q}}\right)$.
In the continuum model, bare current vertex reduces to the usual form
$\Gamma_{x}^{(p)}(\boldsymbol{k},\boldsymbol{k}+\boldsymbol{q})=\frac{1}{2m}(2k_{x}+q_{x})$.

To account for the effects of interactions, the energy flow correlation
function is separated into the free and interaction parts: $\Pi_{xx}^{(p)}=\Pi_{0,xx}^{(p)}+\Pi_{v,xx}^{(p)}$.
The interaction part $\Pi_{v,xx}^{(p)}$ is computed as \citep{Khatibi2020}:

\begin{equation}
\Pi_{v,xx}^{(p)}(q)=-\Pi_{lc,xx}^{(p)}(q)V(q)\Pi_{rc,xx}^{(p)}(q),
\end{equation}
where $V(q)$ represents the particle interaction, and:

\begin{equation}
\Pi_{lc,xx}^{(p)}(q)=\Pi_{rc,xx}^{(p)}(q)=-\frac{1}{\beta N}\sum_{k}\Gamma_{x}^{(p)}(k,k+q)G_{\psi}(k)G_{\psi}(k+q).\label{eq:A particle current correction}
\end{equation}

We now turn to the definition of the energy-current density operator:
\begin{equation}
\boldsymbol{j}^{(E)}(\boldsymbol{r})=\frac{1}{2}\left(\psi^{\dagger}(\boldsymbol{r})\hat{H}(\boldsymbol{r})\hat{v}(\boldsymbol{r})\psi(\boldsymbol{r})+\psi^{\dagger}(\boldsymbol{r})\hat{v}(\boldsymbol{r})\hat{H}(\boldsymbol{r})\psi(\boldsymbol{r})\right).
\end{equation}
Here, $\hat{H}(\boldsymbol{r})$ is the Hamiltonian-density operator
(the local Hamiltonian associated with lattice site $\boldsymbol{r})$,
and $\hat{v}(\boldsymbol{r})=\frac{i}{\hbar}[\hat{H},\hat{\boldsymbol{r}}]$
is the velocity operator. Upon Fourier transformation, the $x$-component
in momentum space is 
\begin{equation}
j_{x}^{(E)}(\boldsymbol{q})=\frac{1}{N}\sum_{\boldsymbol{k}}\left[\frac{1}{2}\left(E(\boldsymbol{k}+\boldsymbol{q})v_{x}(\boldsymbol{k})+E(\boldsymbol{k})v_{x}(\boldsymbol{k}+\boldsymbol{q})\right)\psi^{\dagger}(\boldsymbol{k})\psi(\boldsymbol{k}+\boldsymbol{q})\right],
\end{equation}
with $v_{x}(\boldsymbol{k})=\frac{\partial E(\boldsymbol{k})}{\partial k_{x}}$
the particle velocity.

When evaluating the energy-current correlation function,
\begin{equation}
\Pi_{xx}^{(E)}(x_{2},x_{1})=-\left\langle j_{x}^{(E)}(\boldsymbol{r}_{2},\tau_{2})j_{x}^{(E)}(\boldsymbol{r}_{1},\tau_{1})\right\rangle ,
\end{equation}
the result differs from Eqs. \eqref{eq:A particle current correlation function}
and \eqref{eq:A particle current correction} only in that the particle-current
vertex is replaced by the energy-current vertex $\Gamma_{x}^{(E)}(\boldsymbol{k},\boldsymbol{k}+\boldsymbol{q})=\frac{1}{2}\left(E(\boldsymbol{k}+\boldsymbol{q})v_{x}(\boldsymbol{k})+E(\boldsymbol{k})v_{x}(\boldsymbol{k}+\boldsymbol{q})\right)$,
which directly yields the energy-current correlator..

The heat-current density operator is usually defined by 
\begin{equation}
\boldsymbol{j}^{(Q)}=\boldsymbol{j}^{(E)}-\mu\boldsymbol{j}.
\end{equation}
For practical calculations of heat-current correlation functions,
one may equivalently replace the Hamiltonian density $\hat{H}$ in
$\boldsymbol{j}^{(E)}$ by the grand-canonical Hamiltonian density
$\hat{K}=\hat{H}-\mu\hat{N}$. At the level of vertices in momentum
space, this corresponds to the replacement $E(\boldsymbol{k})\rightarrow\Xi(\boldsymbol{k})\equiv E(\boldsymbol{k})-\mu,$
so that $\Gamma_{x}^{(Q)}(\boldsymbol{k},\boldsymbol{k}+\boldsymbol{q})=\frac{1}{2}\left(\Xi(\boldsymbol{k}+\boldsymbol{q})v_{x}(\boldsymbol{k})+\Xi(\boldsymbol{k})v_{x}(\boldsymbol{k}+\boldsymbol{q})\right)$.

Finally, by substituting $T_{\mathrm{self}}(q)$ into $V_{\mathrm{eff}}$,
and setting $\psi$ to the spinon operator $f$, the spinon heat current
correlation function is thus obtained.

In this appendix, we additionally calculate the spinon particle current-current
correlator $\Pi_{xx}^{(p)}$ and the particle conductivity: 
\begin{equation}
\sigma_{f}(T)=-\lim_{\omega\rightarrow0}\frac{\mathrm{Im}[\Pi_{f}^{(p)}(\boldsymbol{q}=0,\omega)]}{\omega}.\label{eq: A particle conductivity}
\end{equation}
The frequency dependence of the zero-momentum component of $-\mathrm{Im}[\Pi_{f}^{(p)}(\omega+i0^{+})]$
is shown in the Fig. \ref{App spinon conductivity} (a). One again
observes a pronounced Drude peak together with characteristic Fermi-liquid
behavior. The resulting DC conductivity $\sigma_{f}(T)$ scales as
$T^{-3}$, as displayed in Fig. \ref{App spinon conductivity}. In
particular, panel (b) is obtained from the retarded correlator via
NAC, whereas panel (c) is estimated from the lowest Matsubara-frequency
component, $-\mathrm{Im}[\Pi_{f}^{(p)}(i\nu_{1})/i\nu_{1}]$.

Given the $T^{-3}$ scaling of the spinon DC conductivity, together
with the thermal conductivity relation discussed in the main text,
we conclude that the spinon sector itself still satisfies the Wiedemann-Franz
law under one-loop level: $\frac{\kappa_{f}}{\sigma_{f}T}=L_{0,f}$.
\end{widetext}

\begin{figure*}
\begin{centering}
\includegraphics[width=1\linewidth]{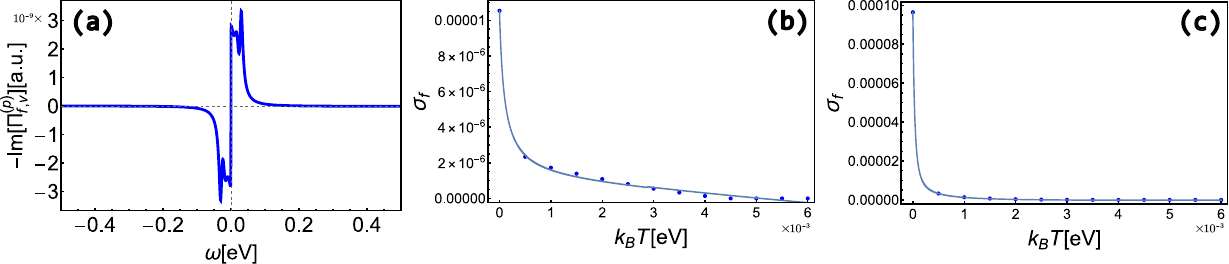}
\par\end{centering}
\caption{(a) Frequency dependence of the imaginary part of the zero-momentum
spinon current-current correlator. (b), (c) Temperature dependence
of the spinon DC conductivity; the data points are obtained from NAC
and from the lowest Matsubara-frequency approximation, respectively.
The solid lines are fits to the function $a_{0}+a_{1}x+b_{1}/x+b_{2}/x^{2}$.}
\label{App spinon conductivity}
\end{figure*}

\onecolumngrid

\section{Thermodynamic potential and Luttinger--Ward Functional of the U(1)
QSL \label{Appendix Thermodynamic potential}}

To present the system's thermodynamic properties from another perspective,
one may compute the thermodynamic potential from which the free energy,
specific heat, and other thermodynamic quantities can be derived.
By definition, the thermodynamic potential is composed of three parts,
$\Omega=\Omega_{f}+\Omega_{X}+\Omega_{\mathrm{LW}}$ where the LW
functional contribution, $\Omega_{\mathrm{LW}}$, is defined in Eq.
\eqref{eq:LW functional}. Each term is expressed as follows \citep{Yanase2005,Blaizot2021,Dupuis2023}:

\begin{align}
\Omega_{f}= & -\frac{1}{\beta N}\sum_{q}\left[\ln\left[(-G_{f}(q))^{2}\right]+2G_{f}(q)\Sigma(q)\right],\nonumber \\
\Omega_{X}= & -\frac{1}{\beta N}\sum_{q}\left[\ln\left[-G_{X}(q)\right]+G_{X}(q)\Pi(q)\right],\nonumber \\
\Omega_{\mathrm{self}}= & \frac{1}{\beta N}\sum_{q}\ln\left[1-2T_{f/X,\mathrm{self},0}(q)\chi_{0}^{f/X}(q)\right],\nonumber \\
\Omega_{\mathrm{bind}}= & \frac{1}{\beta N}\sum_{q}\ln\left[1-2t(\boldsymbol{q})G_{c,0}(q)\right],\nonumber \\
\Omega_{\mathrm{HF}}= & \frac{1}{N}\sum_{ij}2t_{ij}G_{c,0}(x_{j},x_{i}).\label{eq:thermaldynamic potential}
\end{align}

\begin{figure*}
\begin{centering}
\includegraphics[width=0.5\linewidth]{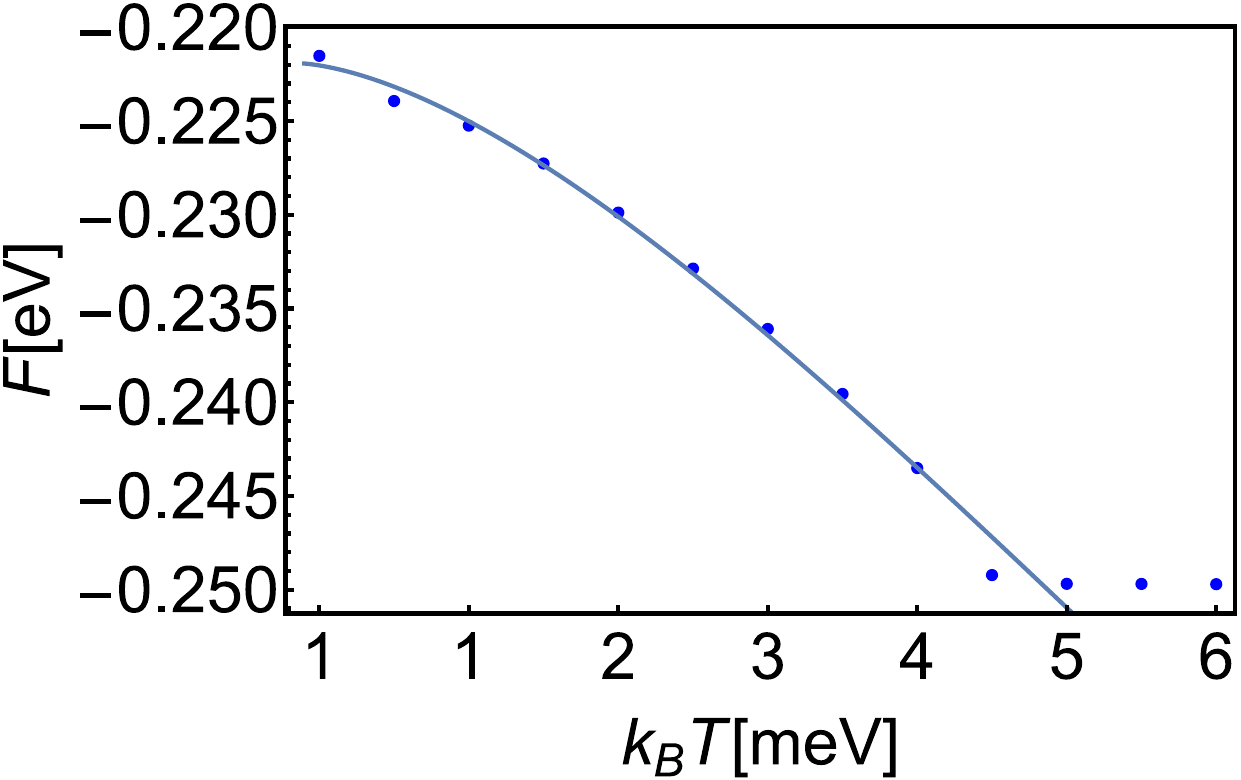} 
\par\end{centering}
\caption{Temperature dependence of the free energy obtained from the thermodynamic
potential defined by LW-functional.}
\label{Free energy} 
\end{figure*}

To overcome difficulties in numerical evaluation of Matsubara frequency
sums, we follow the strategy in \citep{Yanase2005} and write $\Omega=\Omega-\Omega_{e}+\Omega_{e}\approx\widetilde{\Omega}-\widetilde{\Omega}_{e}+\Omega_{e},$where
$\Omega_{e}=-2T\sum_{\boldsymbol{k}}\log\left[1+\exp(-\beta(t(\boldsymbol{k})-\mu))\right]$
is the exact free electron thermodynamic potential, and tildes denote
results from approximate Matsubara sums. Because the interacting Green's
functions share the similar properties as the free-particle Green's
functions at large $\omega_{n}$, this replacement uses the solvable
free electron tail to regularize the problematic high frequency region.
Finally, one computes the free energy $F=\Omega+\mu n$ shown in Fig.
\ref{Free energy}. Additionally, the specific heat $C_{v}=-T\frac{\partial^{2}F}{\partial T^{2}}$
in Fig. \ref{thermal conductivity} (f) is obtained by fitted expression
of free energy data, it also confirms a linear growth of $C_{v}$
with temperature, as expected for a Fermi liquid.

\onecolumngrid

\section{Self-Consistent Equations of Impurity coupled quantum spin liquid
Hamiltonian Under First-Order Born Approximation\label{Appendix Impurity}}

We consider the case of infinite single-impurity scattering processes.
Since the original Hamiltonian does not include the self-energy from
impurity coupling, it must be manually introduced and solved self-consistently,
which is equivalent to the first-order self-consistent Born approximation.
Assuming that the spinon and chargon Green's functions after impurity
coupling take the following forms:

\begin{align}
G(\psi,\psi^{\dagger},i\omega_{n},\sigma)= & \left[\left(\begin{array}{cc}
i\omega_{n}-\epsilon_{d}+h_{2} & 0\\
0 & G^{-1}(f,f^{\dagger},i\omega_{n},\sigma,\boldsymbol{r}=0)
\end{array}\right)-\Sigma_{\mathrm{hybrid}}\right]^{-1},\nonumber \\
G(\phi,\phi^{\dagger},i\nu_{n})= & -\left[\left(\begin{array}{cc}
-(i\nu_{n}+h_{2})^{2}/U_{\mathrm{imp}}+\lambda & 0\\
0 & G^{-1}(X,X^{\dagger},i\nu_{n},\boldsymbol{r}=0)
\end{array}\right)+\Pi_{\mathrm{hybrid}}\right]^{-1}.
\end{align}
Here, $\psi=(a,f)^{T}$ and $\phi=(Y,X)^{T}$ are the composite operator
bases for spinons and chargons, respectively. $\Sigma_{\mathrm{hybrid}}$
and $\Pi_{\mathrm{hybrid}}$ are the self-energies of spinons and
chargons due to their coupling, which are assumed to be frequency-independent.
Specifically, we have:

\begin{align}
\Sigma_{\mathrm{hybrid}}= & \left(\begin{array}{cc}
0 & w\\
w & 0
\end{array}\right)\nonumber \\
\Sigma_{\mathrm{hybrid}}= & \left(\begin{array}{cc}
0 & -u\\
-u & 0
\end{array}\right)
\end{align}
Next, the self-energy equations are derived based on the scattering
processes:

\begin{align}
\Sigma_{\mathrm{hybrid}}= & -VG(\phi,\phi^{\dagger},\tau=0),\nonumber \\
\Pi_{\mathrm{hybrid}}= & -VG(\psi,\psi^{\dagger},i\omega_{n},\sigma,\tau=0).
\end{align}
By expanding these expressions, the self-consistent equations \eqref{eq: SCE u}-
\eqref{eq: SCE w} presented in the main text are obtained. 
\end{document}